\documentclass[12pt]{article}
\usepackage{amsmath}
\usepackage{amssymb}
\usepackage{amstext}
\usepackage{amsfonts,color}
\usepackage{graphicx,epsfig}
\usepackage{slashbox}
\usepackage{indentfirst}
\usepackage[vcentermath]{youngtab}

\def \bea{\begin{eqnarray}}
\def \beq{\begin{equation}}

\def \eea{\end{eqnarray}}
\def \eeq{\end{equation}}

\begin{document}

\baselineskip 20pt

\title{Revisit Spectrum of Baryonium in Heavy Baryon Chiral
Perturbation Theory}
\author{Yue-De Chen$^a$\footnote{Email:
chenyuede@ucas.ac.cn},\
 Cong-Feng Qiao$^{a}$\footnote{Email: qiaocf@ucas.ac.cn},\
 Peng-Nian Shen$^{e,b,c}$\footnote{Email: shenpn@ihep.ac.cn} and
Zhuo-Quan Zeng$^d$\footnote{Email: zengzhuoquan@hotmail.com}\\[0.5cm]
\small $a)$ Department of Physics, University of the Chinese
Academy of Sciences \\ \small YuQuan Road 19A, 100049, Beijing,
China\\
 {\small $b)$ Theoretical Physics Center for Science Facilities
(TPCSF), CAS}\\
{\small YuQuan Road 19B, 100049, Beijing, China}\\
{\small $c)$ Institute of High Energy Physics, CAS, Beijing, 100049,
China}\\
{\small $d)$ College of Physics and Electronic Engineering,
Hainan Normal University, Haikou, 571158, China}\\
{\small $e)$ College of Physics and Technology, Guangxi Normal
University, Guilin 541004, China}\\ }
\date{}
\maketitle

\begin{abstract}
In the framework of the heavy baryon perturbation theory, in which
the two-pion exchange is considered, the physical properties of
heavy-baryon-anti-heavy-baryon systems are revisited. The potentials
between heavy-baryon and anti-heavy-baryon are extracted in a holonomic
form. Based on the extracted potentials, the s-wave
scattering phase shifts and scattering lengths of
$\Lambda_c$-$\bar{\Lambda}_c$ and $\Sigma_c$-$\bar{\Sigma}_c$ are
calculated. From these scattering features, it is found that the
$\Lambda_c$-$\bar{\Lambda}_c$ system can be bound only when
the value of the coupling constant $g_2$ is larger than that
from the decay data of the $\Sigma_c(\Sigma_c^*) \to
\Lambda_c \pi$ process. The binding condition for the
$\Sigma_c$-$\bar{\Sigma}_c$ system is also examined. The binding
possibilities of these systems deduced from the scattering
calculations are also checked by the bound state calculation and the
binding energies are obtained if the system can be really bound. The
binding possibility of the $\Lambda_b$-$\bar{\Lambda}_b$ system is
investigated as well.

\end{abstract}

\section{Introduction}

Charmonium and bottomonium are important objects in
studying strong interactions and structures of hadrons. In the past
decade, many new hadron states with heavy flavors, such as $Y(4260)$,
$Y(4360)$, $Y(4660)$, $Z^{\pm}(4430)$, $Y_b(10890)$, and etc.,
have been found in the $e^+e^-$ annihilation and the $B$ meson
decay experiments by $\mathrm{BABAR}$ and $\mathrm{Belle}$ \cite{belle1},
in particular, $Z_c(3900)$ has been observed in the $e^+e^-$ annihilation
by $\mathrm{BESIII}$ and $\mathrm{Belle}$\cite{bes,bel} recently.
These new findings have attracted much attention on the structure of
hadron all over the particle and nuclear physics
societies \cite{N.Bram,xiangL, guo, jian-rong zhang}. However, some of
the states cannot be identified as a conventional quarkonium with heavy
flavor, such as the charmonium or its excited state because of their
abnormal quantum numbers, masses, decay modes and corresponding branching
ratios in experiments. Therefore, as mentioned in our previous paper \cite{chen},
to explain the peculiar data, many postulates for their structures
have been proposed, but up to now, no definite
conclusions could be drawn yet.

One of the striking pictures among the postulates is the baryonium
with heavy flavor. In the extended heavy baryonium picture used in
our previous paper \cite{chen, Qiao}, an approximate SU(2) symmetry
between $\Lambda_c$ and $\Sigma^0_c$ is assumed, and $\Lambda_c$ and
$\Sigma^0_c$ are taken as the basis vectors in the two-dimensional
"C-spin" representation, which is analogues to the isospin in the
nucleon doublet case. Apparently, these basis vectors can form a
"C-spin" triplet and a "C-spin" singlet \cite{Qiao}. The key point
is to verify if a heavy baryon and a heavy antibaryon can really
form a bound state, the baryonium, dynamically. The simplest way to
achieve this aim is extracting a potential between heavy baryons by
using a theory, for instance the so-called heavy baryon chiral
perturbation theory (HBCPT) which can effectively provide a good
description for the heavy baryon, and then solving the
Schr\"{o}dinger equation for the energy eigenvalue and consequently
the mass spectrum.

In fact, in our previous investigation \cite{chen} we have studied the
possibility of forming a heavy baryonium by using HBCPT. The result showed
that there might exist a heavy baryonium as long as the adopted values
of the coupling constant at the baryon-Goldstone-boson vertex and the
cutoff parameter $\Lambda$ are, respectively, in the special ranges,
although the result is very sensitive to such values. Apparently, the
strong parameter dependence is undesirable. Such a dependence might
come from the inappropriate approximation in deriving potentials, for
instance the premature truncation to the term with $1/r^{5/2}$ in the
asymptotic expressions of the potentials expanded in $\lambda$
\cite{chen,richardson}. This is because that the contribution from
the two-pion-exchange potential is short ranged, but the expansion
of the potential
function in $\lambda$ requires a relatively larger $r$. Although the
dominant contribution of the $\lambda$ integral comes from small
$\lambda$ values, the expansion converges extremely slow. Moreover,
whether the physical value of the coupling constant, which can be
extracted from relevant decay data, supports the existence of a heavy
baryonium is still questionable and this problem should further be
investigated carefully.

In this paper, we first re-derive the potential between
$\Lambda_c(\Sigma_c)$ and $\bar{\Lambda}_c(\bar{\Sigma}_c)$ in a
holonomic form rather than a truncated expansion in Ref.\cite{chen}.
Then, we study the $\Lambda_c-\bar{\Lambda}_c$ ($\Sigma_c-\bar{\Sigma}_c$)
scattering to get scattering characters, in particular those closely
related to its binding feature. Based on the enlightenment from scattering
information, we finally calculate the binding behavior of the system to
confirm whether a heavy baryonium really exists. The paper is organized
as follows. In Section 2, the formalism of
HBCPT is briefly recalled. The two body interaction potentials in
the $\Lambda_c^+$-$\bar{\Lambda}_c^+$ and
$\Sigma_c^0$-$\bar{\Sigma}_c^0$ systems are given in Section 3. In
Section 4, the numerical results for the scattering information
and the mass spectra of possible heavy baryonia are presented.
And the summary is given in Section 5.

\section{A brief introduction to HBCPT}

As commonly adopted, symbol $q_1q_2Q$, where $q_{1(2)}$ represents
the light quark, and $Q$ denotes the heavy quark, describes a heavy
baryon which contains one heavy quark and two light quarks. Assuming
that two light quarks form a pair of diquark, then in the flavor space,
these three quarks can form a symmetric sextet and an antisymmetric triplet,
i.e. $3 \otimes 3 = 6 \oplus \bar{3}$. Because the
wave function of the hadron in the color space is totally antisymmetric,
the wave function in the direct product space of orbit, flavor and spin
must be symmetric. Consequently, for a ground state hadron, the wave function
in the flavor and spin spaces should be symmetric since the orbital wave
function is symmetric. For the light quark pair, we use Young table
$\Yvcentermath1\young(12)^F_{q_1,q_2}$ and
$\Yvcentermath1\young(1,2)^F_{q_1,q_2}$ to denote the symmetric sextet and
antisymmetric triplet in the flavor space, respectively, and
$\Yvcentermath1\young(12)^S_{q_1,q_2}$ and $\Yvcentermath1\young(1,2)^S_{q_1,q_2}$
to represent the triplet and singlet in the spin space, respectively.
Coupling these wave functions of a diquark to that of a heavy quark,
denoted by $\Yvcentermath1\young(Q)^F\times \Yvcentermath1\young(Q)^S$, we have
\begin{eqnarray}
&&\left[\left(\Yvcentermath1\young(12)^F_{q_1,q_2} \times
\Yvcentermath1\young(12)^S_{q_1,q_2}\right) \oplus
\left(\Yvcentermath1\young(1,2)^F_{q_1,q_2} \times
\Yvcentermath1\young(1,2)^S_{q_1,q_2}\right)\right] \times
\left(\Yvcentermath1\young(Q)^F\times \Yvcentermath1\young(Q)^S\right) \nonumber\\
& = & \Yvcentermath1\young(12)^F_{q_1,q_2} \times
\Yvcentermath1\young(Q)^F\times
\left(\Yvcentermath1\young(12Q)^S_{q_1,q_2,Q} \oplus
\Yvcentermath1\young(12,Q)^S_{q_1,q_2,Q}\right) \oplus
\left(\Yvcentermath1\young(1,2)^F_{q_1,q_2} \times
\Yvcentermath1\young(Q)^F \times
\Yvcentermath1\young(Q)^S\right).\label{FS}
\end{eqnarray}
Equation (\ref{FS}) suggests that the sextet $\textbf{6}$ has
spin-$\frac{1}{2}$ and spin-$\frac{3}{2}$ states, while the triplet
$\bar{\textbf{3}}$ has only spin-$\frac{1}{2}$ states. Writing them
explicitly in the matrix form, we have
\begin{equation}
B_6=\left(\begin{array}{ccc}\Sigma_c^{++}& \frac{1}
{\sqrt{2}}\Sigma_c^{+} & \frac{1}{\sqrt{2}}\Xi_c^{'+}\\
\frac{1}{\sqrt{2}}\Sigma_c^{+} & \Sigma_c^0 &
\frac{1}{\sqrt{2}}\Xi_c^{'0}\\
\frac{1}{\sqrt{2}}\Xi_c^{'+} & \frac{1}{\sqrt{2}}\Xi_c^{'0} &
\Omega_c^0\end{array}\right)\; \label{B6}
\end{equation}
and
\begin{equation}
B_{\bar{3}}=\left(\begin{array}{ccc}0& \Lambda_c & \Xi_c^{+}\\
-\Lambda_c & 0 &
\Xi_c^{-}\\
-\Xi_c^{+} & -\Xi_c^{-} & 0\end{array}\right)\; \label{B3}
\end{equation}
for the sextet and triplet of the charmed heavy baryon,
respectively, and the same form of Eq.(\ref{B6}) for the
spin-$\frac{3}{2}$ $B_6^*$ multiple.
These forms are also applicable to the bottomed heavy baryon multiple
by substituting $c$ with $b$.

On the other hand, in terms of the chiral perturbation theory, one has the
leading order vector and axial vector fields in $f_{\pi}$  \cite{A.Manohar, M.Wise, chen}
\begin{equation}
V_\mu=\frac{1}{f_\pi^2}M\partial_\mu M\; ,
\end{equation}
\begin{equation} A_\mu=-\frac{1}{f_\pi}\partial_\mu M\; ,
\end{equation}
where $M$ is the Goldstone boson matrix
\beq
 M =
\left(\begin{array}{ccc} \frac{1}{\sqrt{2}} \pi^0 +
\frac{1}{\sqrt{6}} \eta &\phantom{+} \pi^+
& \phantom{+} K^+ \\
\phantom{+}  \pi^- &  -\frac{1}{\sqrt{2}} \pi^0 + \frac{1}{\sqrt{6}}
\eta & \phantom{+} K^0\\
\phantom{+} K^- &  \phantom{+} \bar{K}^0 & \phantom{+} -
\frac{2}{\sqrt{6}} \eta  \end{array}\right) \; . \eeq

Then, the general form of the chiral-invariant Lagrangian can be written as
\cite{T.M.Yan}
\begin{eqnarray}
\mathcal{L}& =
&\frac{1}{2}tr[\bar{B}_{\bar{3}}(iD\!\!\!/-M_{\bar{3}})
B_{\bar{3}}]+tr[\bar{B}_6(iD\!\!\!/-M_6)B_6]\nonumber\\
&+&tr[\bar{B}_6^{*\mu}[-g_{\mu\nu}(iD\!\!\!/-M_6^{*})+i(\gamma_\mu
D_\nu+\gamma_\nu
D_\mu)-\gamma_\mu(iD\!\!\!/+M_6^{*})\gamma_\nu]B_6^{*\nu}]\nonumber\\
&+&g_1tr(\bar{B}_6\gamma_{\mu}\gamma_5A^{\mu}B_6)+g_2tr(\bar{B}_6
\gamma_{\mu}\gamma_5A^{\mu}B_{\bar{3}})+ h.c.\nonumber\\
&+&g_3tr(\bar{B}_{6{\mu}}^*A^{\mu}B_6)+ h.c. + g_4
tr(\bar{B}_{6{\mu}}^*A^{\mu}B_{\bar{3}}) + h.c. \nonumber\\
&+&g_5tr(\bar{B}_6^{\nu*}\gamma_{\mu}\gamma_5A^{\mu}
B_{6\nu}^*)+g_6tr(\bar{B}_{\bar{3}}\gamma_{\mu}\gamma_5A^{\mu}B_{\bar{3}})\;,
\end{eqnarray}
where the chiral covariant derivative $D_\mu$ satisfies
\begin{equation}
D_\mu B_6 = \partial_\mu B_6 + V_\mu B_6 + B_6 V_\mu ^T \;,
\end{equation}
\begin{equation}
D_\mu B_{\bar{3}} = \partial_\mu B_{\bar{3}} + V_\mu B_{\bar{3}} +
B_{\bar{3}} V_\mu ^T \;.
\end{equation}
According to the heavy quark symmetry, six coupling constants approximately obey
following relations:
\begin{eqnarray}
g_1 = \frac{2\sqrt{3}}{3}g_3 = -\frac{2}{3}g_5 \; , \;\;\;\;  g_2 =
-\frac{\sqrt{3}}{3}g_4 \; ,\;\;\;\;  g_6 = 0 \; , \label{couplings}
\end{eqnarray}
thus, we have only two free parameters $g_1$ and $g_2$ in the numerical
calculation \cite{T.M.Yan}.

\section{The formulation for fwo body scattering potential}

\begin{figure}[t,m]
\begin{center}
\parbox{.45\textwidth}{\epsfysize=2.5cm \epsffile{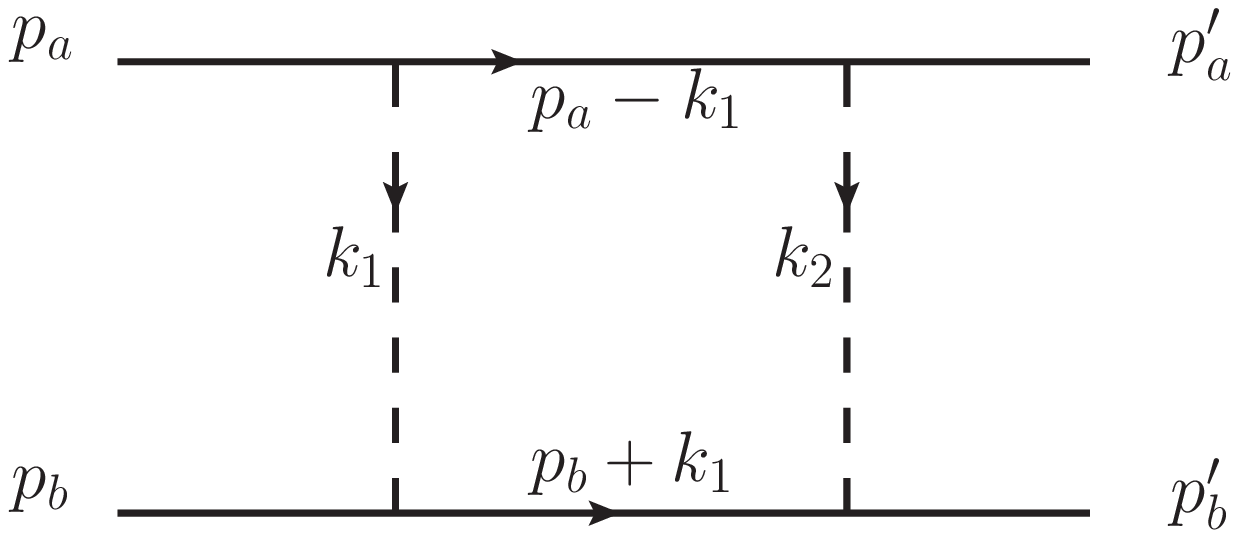}}
\parbox{.35\textwidth}{\epsfysize=2.5cm \epsffile{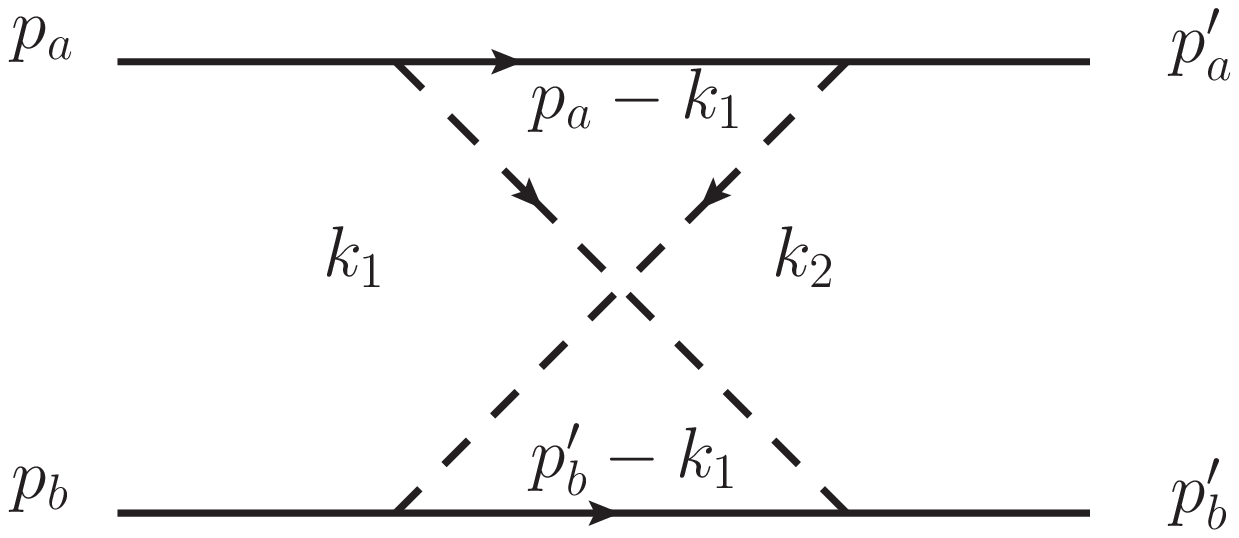}}\hfill
\hspace*{3.5cm}\parbox{0.45\textwidth}{\hspace{.5cm}
(a)}
\parbox{0.35\textwidth}{\hspace{0.6cm}(b)}\\
\parbox{.45\textwidth}{\epsfysize=2.5cm \epsffile{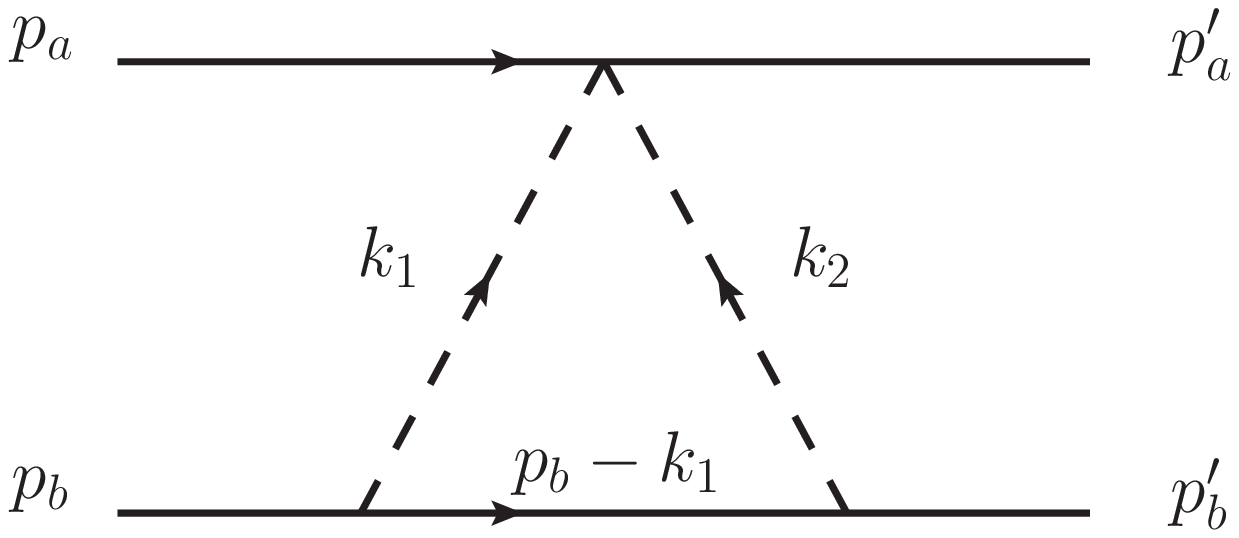}}
\parbox{.35\textwidth}{\epsfysize=2.5cm \epsffile{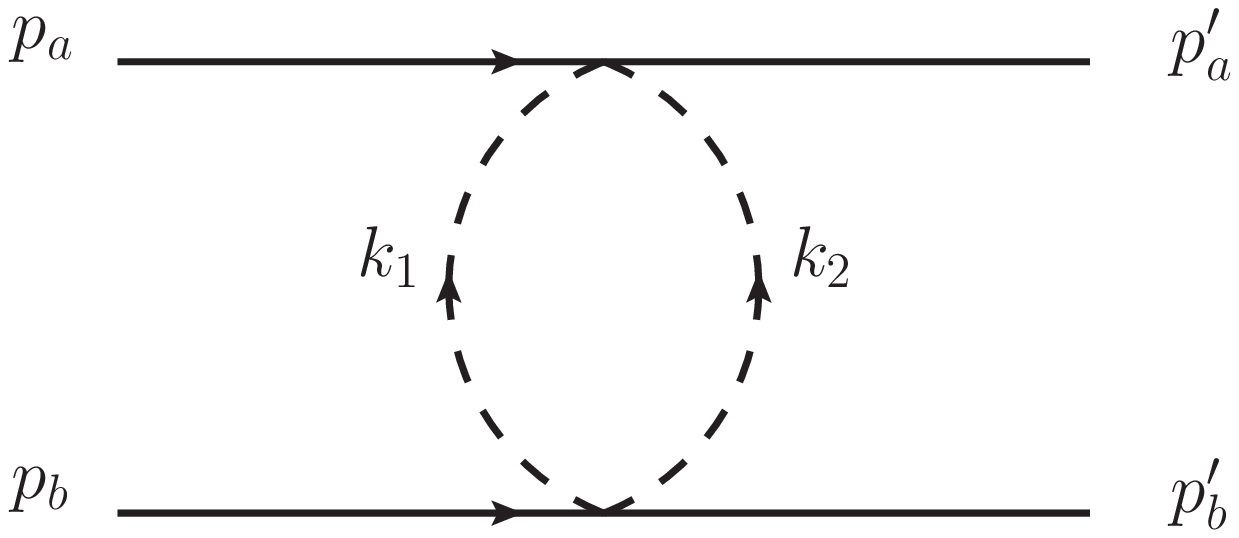}}\hfill
\hspace*{3.5cm}\parbox{0.45\textwidth}{\hspace{.5cm}
(c)}
\parbox{0.35\textwidth}{\hspace{0.6cm}(d)}\\
 \caption{$2\pi$-exchange diagrams: (a) box diagram, (b) crossed diagram,
 (c) triangle diagram, (d) two-pion loop diagram. }
  \label{fig:twopi}
  \end{center}
\end{figure}

To derive the two body scattering kernel and further the potential,
as carried out in Ref.\cite{chen}, we follow the technique in
Refs.\cite{Th.Rijken1, Th.Rijken2}. We first write down the scattering
amplitude to get the interaction kernel, and then, make the non-relativistic
reduction. Further making Fourier transformation, we obtain the potential
in the configuration space. Then, acting the operators onto a considered
channel, we finally obtain the potential for such a particular system.
Again, in this continuation paper, we calculate the potentials in four
$2\pi$-exchange diagrams shown in Fig.\ref{fig:twopi}.

In the center of mass system (CMS), we define
\begin{eqnarray}
&\textbf{p}_a &= -\textbf{p}_b = \textbf{p}, \nonumber \\
&\textbf{p}'_a &= -\textbf{p}'_b = \textbf{p}' \nonumber \\
&P  &= p_a + p_b = (E_a + E_b\;,0 ) = (E\;,0), \nonumber \\
&P' &= p_a'+ p_b' = (E'_a + E'_b\;,0 ) = (E'\;,0)\;, \nonumber  \\
&p  &= \frac{1}{2}(p_a - p_b) = (0\;,\textbf{p})\;,  \nonumber \\
&p' &= \frac{1}{2}(p'_a - p'_b) = (0\;,\textbf{p}') \nonumber \;,
\end{eqnarray}
as the three-momenta of the initial and final states, the total
four-momenta of the initial and final states, and the relative
four-momenta of the initial and final states, respectively.

For the $\Lambda_c - \bar{\Lambda}_c$ interaction, we first calculate
the potential in the box diagram. Following the prescription in
Refs.\cite{Th.Rijken1, Th.Rijken2}, we obtain the potential in the
configuration space (detailed calculation can be found in Ref.\cite{chen})
\begin{eqnarray}
V_B (r_1,\; r_2)=-\left(\frac{g_4^4}{f_{\pi}^4}\right)\int\int
\frac{d^3\textbf{k}_1d^3\textbf{k}_2}{(2\pi)^6}
\frac{\mathcal{O}_1(\textbf{k}_1,\textbf{k}_2)
e^{i\textbf{k}_1\textbf{r}_1}e^{i\textbf{k}_2\textbf{r}_2}
f(\textbf{k}_1^2)f(\textbf{k}_2^2)}
{2E_{\textbf{k}_1}E_{\textbf{k}_2}
(E_{\textbf{k}_1}+\Delta)(E_{\textbf{k}_2}
+\Delta)(E_{\textbf{k}_1}+E_{\textbf{k}_2})}\; . \label{cp1}
\end{eqnarray}
As commonly used, we take a Gaussian form for the form factor $f(\textbf{k}_i^2)$
which regulates the integral. Further using the integral factorization technique,
we get the non-local central potential
\begin{eqnarray}
V_B(r_1\;,r_2) = &-&\left(\frac{g_4^4}{f_{\pi}^4}\right) \frac{1}{\pi}
\mathcal{O}_1(\textbf{k}_1\;,\textbf{k}_2)\left[\int_0^\infty
\frac{d\lambda}{\Delta^2 + \lambda^2}
F(\lambda,\;r_1)F(\lambda,\;r_2)\right.\nonumber\\&&\left.
-\frac{2\Delta}{\pi^2} \int_0^\infty\frac{d\lambda}{\Delta^2 +
\lambda^2} F(\lambda,\;r_1) \int_0^\infty\frac{d\lambda}{\Delta^2 +
\lambda^2} F(\lambda,\;r_2)\right],
\end{eqnarray}
where $\Delta = M_{\Sigma_c^{\prime}}-M_{\Lambda_c}$ with $\Sigma_c^{\prime}$
being either $\Sigma^{+}_c$ or $\Sigma^{+*}_c$ as the intermediate state in
the $\Lambda^{+}_c - \bar{\Lambda}^+_c$ interaction, and the form of
$F(\lambda,\;r)$ can be found in Ref.\cite{chen}.

In the same way, we can calculate potential in the crossed diagram and get the non-local potential
\begin{eqnarray}
V_C(r_1\;,r_2) = - \left(\frac{g_4^4}{f_{\pi}^4}\right)\frac{1}{\pi}
\mathcal{O}_1(\textbf{k}_1\;,\textbf{k}_2)\int_0^\infty d\lambda
\frac{\Delta^2 - \lambda^2}{(\Delta^2 + \lambda^2)^2}
F(\lambda,\;r_1)F(\lambda,\;r_2).
\end{eqnarray}
Similarly, we obtain the non-local potential in the triangle diagram
\begin{equation}
V_{triangle}(r_1,r_2) =
\frac{g_4^2}{2f_\pi^4}\int\int\frac{d^3\textbf{k}_1d^3\textbf{k}_2}{(2\pi)^6}
\frac{\mathcal{O}_2(\textbf{k}_1,\textbf{k}_2) (E_{\textbf{k}_1} +
E_{\textbf{k}_2})
e^{i\textbf{k}_1\textbf{r}_1}e^{i\textbf{k}_2\textbf{r}_2}
f(\textbf{k}_1^2)f(\textbf{k}_2^2)}
{E_{\textbf{k}_1}E_{\textbf{k}_2}(E_{\textbf{k}_1} + \Delta)
(E_{\textbf{k}_2} + \Delta)}\;,\label{tri11}
\end{equation}
and potential in the $2\pi$-loop diagram
\begin{equation}
V_{2\pi-loop}(r_1, r_2) = \frac{1}{16
f_\pi^4}\int\int\frac{d^3\textbf{k}_1d^3\textbf{k}_2}{(2\pi)^6}
e^{i\textbf{k}_1\textbf{r}_1}e^{i\textbf{k}_2\textbf{r}_2}
f(\textbf{k}_1^2)f(\textbf{k}_2^2) A\;, \label{twopair1}
\end{equation}
where $E_{{\bf k}_i}=\sqrt{{\bf k}_i^2+m^2}$ and $A=-\frac{1}{2
E_{\textbf{k}_1}}-\frac{1}{2 E_{\textbf{k}_2}}
+\frac{2}{E_{\textbf{k}_1}+E_{\textbf{k}_2}}$.  In the above
potentials, the operators $\mathcal{O}_i(\textbf{k}_1,\textbf{k}_2)$
with $i=1,2$ come from the non-relativistic reduction for the
interactive vertices. Their general forms are
\begin{eqnarray}
\mathcal{O}_1(\textbf{k}_1,\; \textbf{k}_2)
&=&c_1(\textbf{k}_1\cdot\textbf{k}_2)^2+
c_2(\boldsymbol{\sigma}_1\cdot\textbf{k}_1
\times\textbf{k}_2)(\boldsymbol{\sigma}_2\cdot\textbf{k}_1
\times\textbf{k}_2)\; ,
\label{rdo}\\
\mathcal{O}_2(\textbf{k}_1,\; \textbf{k}_2)&=&\left( \textbf{k}_1
\cdot \textbf{k}_2\right)\; .\label{rdo2}
\end{eqnarray}
Note that in the right side of Eq.(\ref{rdo}), the first term
will generate a central potential and the second term will produce
a spin-spin potential and a tensor potential.

Finally, acting $\mathcal{O}_i(\textbf{k}_1,\; \textbf{k}_2)$ onto the
concerned channel, making local approximation, and working out detailed
derivation, for the box diagram, we obtain a central potential
\begin{eqnarray}
V_{BC}(r) = &-&\left[\frac{1}{\pi}\int_0^\infty
\frac{d\lambda}{\Delta^2 + \lambda^2}F_C(\lambda,
r)- \frac{4\Delta}{\pi^2
r^2}\left(\int_0^\infty \frac{d\lambda}{\Delta^2 + \lambda^2}
F'(\lambda,\; r)\right)^2 \right.\nonumber\\
&& \left.- \frac{2\Delta}{\pi^2}\left(\int_0^\infty
\frac{d\lambda}{\Delta^2 + \lambda^2}F''(\lambda,\;
r)\right)^2\right],
\end{eqnarray}
a spin-spin potential
%
\begin{eqnarray}
V_{BS}(r)(\boldsymbol{\sigma}_1\cdot \boldsymbol{\sigma}_2) =
&-&\left[\frac{2}{3\pi}\int_0^\infty \frac{d\lambda}{\Delta^2 +
\lambda^2}F_S(\lambda, r) \right.\nonumber\\&&
 - \frac{4\Delta}{3\pi^2}
\left(\frac{1}{r^2}\int_0^\infty\frac{d\lambda}{\Delta^2 +
\lambda^2} F'(\lambda,\;r) \int_0^\infty\frac{d\beta}{\Delta^2 +
\beta^2} F'(\beta,\; r) \right.\nonumber\\&&\left.\left.
 + \frac{2}{r}\int_0^\infty\frac{d\lambda}{\Delta^2 + \lambda^2}
F'(\lambda,\;r) \int_0^\infty\frac{d\beta}{\Delta^2 + \beta^2}
F''(\beta,\; r)\right)\right](\boldsymbol{\sigma}_1\cdot
\boldsymbol{\sigma}_2),
\end{eqnarray}
and a tensor potential
\begin{eqnarray}
V_{BT}(r) S_{12} = &-&\left[\frac{2}{3\pi}
\int_0^\infty\frac{d\lambda}{\Delta^2 + \lambda^2}F_T(\lambda, r)
\right.\nonumber\\&&
 -\frac{4\Delta}{3\pi^2}
\left(\frac{1}{r^2}\int_0^\infty\frac{d\lambda}{\Delta^2 +
\lambda^2} F'(\lambda,\;r) \int_0^\infty\frac{d\beta}{\Delta^2 +
\beta^2} F'(\beta,\; r) \right.\nonumber\\&&\left.\left.
 - \frac{1}{r}\int_0^\infty\frac{d\lambda}{\Delta^2 + \lambda^2}
F'(\lambda,\;r) \int_0^\infty\frac{d\beta}{\Delta^2 + \beta^2}
F''(\beta,\; r)\right)\right] S_{12}.
\end{eqnarray}
For the crossed diagram, the central, spin-spin and tensor
potentials are
\begin{eqnarray}
V_{CC}(r) = -\left[\frac{1}{\pi}\int_0^\infty d\lambda
\frac{\Delta^2 - \lambda^2}{(\Delta^2 + \lambda^2)^2} F_C(\lambda,
r)\right],
\end{eqnarray}
\begin{eqnarray}
V_{CS}(r) = -\left[\frac{1}{\pi}\int_0^\infty d\lambda
\frac{\Delta^2 - \lambda^2}{(\Delta^2 + \lambda^2)^2}F_S(\lambda,
r)\right],
\end{eqnarray}
and
\begin{eqnarray}
V_{CT}(r) = -\left[\frac{1}{\pi}\int_0^\infty d\lambda
\frac{\Delta^2 - \lambda^2}{(\Delta^2 + \lambda^2)^2}F_T(\lambda,
r)\right],
\end{eqnarray}
respectively, where $F_C$, $F_S$, and $F_T$ are
\begin{eqnarray}
F_C(\lambda,r)&=&\frac{2}{r^2}F'(\lambda,r)F'(\lambda,r)+F''(\lambda,r)F''(\lambda,r),\\
F_S(\lambda,r)&=&\frac{F'(\lambda,r)}{r} \left (
\frac{F'(\lambda,r)}{r} +2F''(\lambda,r)\right),\\
F_T(\lambda,r)&=&\frac{F'(\lambda,r)}{r} \left (
\frac{F'(\lambda,r)}{r} -F''(\lambda,r)\right),
\end{eqnarray}
respectively.

For the triangle diagram, at the order of
$\mathcal{O}(\frac{1}{M_H})$, we have only a
central potential
\begin{eqnarray}
V_{TC}(r) = \frac{4 \Delta}{\pi^2} \int_0^\infty d\lambda
\frac{\lambda^2}{\Delta^2 + \lambda^2} F'(\lambda,\; r)
\int_0^\infty \frac{d \lambda}{\Delta^2 + \lambda^2}F'(\lambda,\;
r).
\end{eqnarray}
Similarly, for the $2\pi$-loop diagram, only a central potential
contributes
\begin{eqnarray}
V_{2\pi-loop}(r) = - \frac{2}{\pi}\left[\int_0^\infty d\lambda F
(\lambda\;,r)
\left(\frac{\Lambda^3}{8\pi^{3/2}}\exp(-\frac{1}{4}\Lambda^2 r^2) -
2\lambda^2 F(\lambda\;,r)\right)\right].
\end{eqnarray}
Summing up all the potentials, we eventually obtain the
two-pion-exchange potential for the heavy-baryon-anti-heavy-baryon
interaction
\begin{equation}
V(r) = V_C(r) + V_S(r) \boldsymbol{\sigma}_1\cdot
\boldsymbol{\sigma}_2 + V_T(r) {\bf{S}}_{12}\;,
\end{equation}
where $V_C(r)$, $V_S(r)$ and $V_T(r)$ are the radial parts of the central,
spin-spin and tensor potentials, respectively. From above potential forms,
we see that the longest range of the obtained potentials is, as expected,
that of the two-pion-exchange, because they have a quadratic product of
$F(\lambda, r)$ (or derivatives), and thus have their longest range terms
proportional to $exp(-2mr)$. In addition, we would point out that because
the kernels in different channel are the same except the coefficients,
for simplicity, we can derive the kernel itself first, and then add the
coefficient later for the particular system.

Now, we go to specific systems.

\subsection{$\Lambda_c^+$-$\bar{\Lambda}_c^+$ potential}

In the $\Lambda_c^+$-$\bar{\Lambda}_c^+$ interaction, we assume that
both $\Sigma_c^+$ and $\Sigma_c^{+*}$ could be the intermediate state.
The Lagrangian for spin-$\frac{1}{2}$ $\Sigma_c$-$\pi$-$\Lambda_c$ interaction reads
\begin{eqnarray}
\mathcal{L}_{\Sigma_c-\pi-\Lambda_c} &=& -\frac{g_2}{f_\pi}~
\bar{\Sigma}^{++,~+,~0}_c ~\gamma^\mu\gamma_5~\partial_\mu
\pi^{+,~0,~-} ~\Lambda^{+}_c +h.c.,
\label{eq:sigpilam}
\end{eqnarray}
where the strong coupling constant $g_2$ can be extracted from
$\Sigma_c^{++}\rightarrow \Lambda_c^+ + \pi^+$ decay process (see
Fig.\ref{g2}) by
\begin{equation}
\Gamma = \frac{g_2^2 |\textbf{k}|}{8f_\pi^2
M_{\Sigma_c^{++}}^2}(M_{\Sigma_c^{++}}^2 + M_{\Lambda_c^{+}}^2)
\left[(M_{\Sigma_c^{++}} - M_{\Lambda_c^{+}})^2-m_\pi^2\right],
\label{width1}
\end{equation}
where $|\textbf{k}|=94$MeV is the momentum of the pion in the
$\Sigma_c^{++}$ rest frame, $f_{\pi}=0.132$GeV, $M_{\Sigma_C^{++}}=0.245$GeV,
$M_{\Lambda_{c}^+}=0.229$GeV, and $\Gamma=2.23\pm 0.30$MeV \cite{PDG}. The
resultant phenomenological coupling constant is $g_2 = 0.5 \pm 0.07 $.
\begin{figure}[htbp]
\begin{center}
\parbox{.45\textwidth}{\epsfysize=2.5cm \epsffile{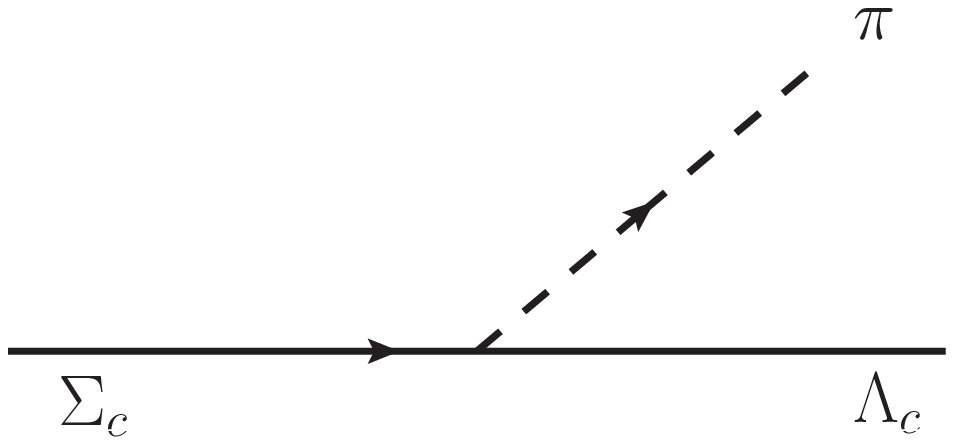}}
\scalebox{1.0}{\hspace{-1.5cm} \begin{small} = i$\mathcal{A} =
\frac{g_{2}}{f_\pi}\bar{u}(p) i\gamma^\mu\gamma_5 k_\mu u(q)$ \end{small}}
 \caption{Vertex of the $\Sigma_c^*-\pi-\Lambda_c$ interaction
for extracting $g_2$.}
  \label{g2}
  \end{center}
\end{figure}

\noindent
Based on the Lagrangian in Eq.(\ref{eq:sigpilam}), we have an explicit
form of $\mathcal{O}_1(\textbf{k}_1,\textbf{k}_2)$ for both box and crossed diagrams
\begin{eqnarray}
\mathcal{O}_1(\textbf{k}_1,\textbf{k}_2)
& = &(\textbf{k}_1\cdot\textbf{k}_2)^2+
(\boldsymbol{\sigma}_1\cdot\textbf{k}_1
\times\textbf{k}_2)(\boldsymbol{\sigma}_2\cdot\textbf{k}_1
\times\textbf{k}_2)\;.
\end{eqnarray}
It leads to a $\Lambda_c^+$-$\bar{\Lambda}_c^+$ potential, caused by the
$2\pi$-exchange with $\Sigma_c$ as the intermediate state,
\begin{eqnarray}
V_{1\Lambda_c^+\bar{\Lambda}_c^+} (r)&=&
\frac{g_2^4}{f_\pi^4}\left[V_{BC}(r) + V_{CC}(r) +
V_{TC}(r)\right] +\frac{1}{f_\pi^4} V_{2\pi-loop}(r)\nonumber\\
&+&\frac{g_2^4}{f_\pi^4}\left[V_{BS}(r) +
V_{CS}(r)\right](\boldsymbol{\sigma}_1\cdot \boldsymbol{\sigma}_2) +
\frac{g_2^4}{f_\pi^4}\left[V_{BT}(r) + V_{CT}(r)\right]S_{12}.
\end{eqnarray}

The Lagrangian for spin-$\frac{3}{2}$ $\Sigma^*_c$-$\pi$-$\Lambda_c$
interaction can be written as
\begin{eqnarray}
\mathcal{L}_{\Sigma^*_c-\pi-\Lambda_c} &=& -\frac{g_4}{f_\pi}
~(\bar{\Sigma}_c^{++,~+,~0})^{*\mu}~ \partial_\mu \pi^{+,~0,~-}
~\Lambda_c^+ +h.c.\; .
\label{eq:sig*pilam}
\end{eqnarray}
Similarly, the coupling constant $g_4$ can be extracted from the
$\Sigma_c^*\rightarrow \Lambda_c + \pi$ decay process by
\begin{equation}
\Gamma = \frac{g_4^2|\textbf{k}|M_{\Sigma^*_c}^2}{96f_\pi^2}
\left[(1-\frac{M_{\Lambda_c}}{M_{\Sigma^*_c}})^2 -
\frac{m_\pi^2}{M_{\Sigma^*_c}^2}\right] \left[(1 +
\frac{M_{\Lambda_c}}{M_{\Sigma^*_c}})^2 -
\frac{m_\pi^2}{M_{\Sigma^*_c}^2}\right]^2,\label{width2}
\end{equation}
with $|\textbf{k}|=180$MeV being the momentum of the pion in the rest
frame of $\Sigma_c^{++*}$, $M_{\Sigma_C^{++*}}=0.252$GeV,
$M_{\Lambda_{c}^{0*}}=0.229$GeV, and $\Gamma=14.9\pm 1.9$MeV \cite{PDG}.
The obtained coupling constant is $g_4 = 0.57 \pm 0.07$. Apparently,
resultant $g_4$ and $g_2$ are not agreed with the
symmetry relation shown in Eq.(\ref{couplings}). This implies
that the heavy quark symmetry is broken.
\begin{figure}[htbp]
\begin{center}
\parbox{.45\textwidth}{\epsfysize=2.5cm \epsffile{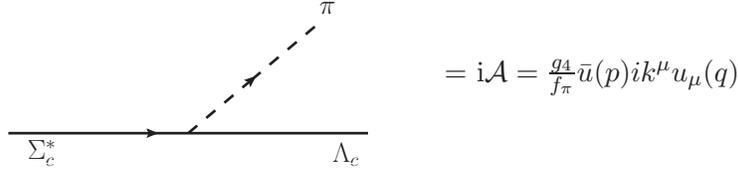}}
\scalebox{1.0}{\hspace{-1.5cm} \begin{small} = i$\mathcal{A} =
\frac{g_4}{f_\pi}\bar{u}(p) ik^\mu u_\mu (q)$ \end{small}}
 \caption{Vertex of the $\Sigma_c^*-\pi-\Lambda_c$ interaction
for extracting $g_4$.}
  \label{g2}
  \end{center}
\end{figure}

Similar to the above case, using Lagrangian in Eq.(\ref{eq:sig*pilam})
we can explicitly write out $\mathcal{O}_1(\textbf{k}_1,\textbf{k}_2)$ for the box
diagram as
\begin{eqnarray}
\mathcal{O}_1(\textbf{k}_1,\textbf{k}_2)
& = &\frac{4}{9}(\textbf{k}_1\cdot\textbf{k}_2)^2-
\frac{1}{9}(\boldsymbol{\sigma}_1\cdot\textbf{k}_1
\times\textbf{k}_2)(\boldsymbol{\sigma}_2\cdot\textbf{k}_1
\times\textbf{k}_2)\; ,
\end{eqnarray}
and for the crossed diagram as
%
\begin{eqnarray}
\mathcal{O}_1(\textbf{k}_1,\textbf{k}_2)
& = &\frac{4}{9}(\textbf{k}_1\cdot\textbf{k}_2)^2 +
\frac{1}{9}(\boldsymbol{\sigma}_1\cdot\textbf{k}_1
\times\textbf{k}_2)(\boldsymbol{\sigma}_2\cdot\textbf{k}_1
\times\textbf{k}_2)\; .
\end{eqnarray}
These lead to a $\Lambda_c^+$-$\bar{\Lambda}_c^+$ potential, caused
by the $2\pi$-exchange with $\Sigma_c^*$ as the intermediate state,
\begin{eqnarray}
V_{2\Lambda_c^+\bar{\Lambda}_c^+} (r)&=&
\frac{4g_4^4}{9f_\pi^4}\left[V_{BC}(r) + V_{CC}(r)\right] +
\frac{2 g_4^2}{3 f_\pi^2}V_{TC}(r)\nonumber\\
&+&\frac{g_4^4}{9 f_\pi^4}\left[- V_{BS}(r) +
V_{CS}(r)\right](\boldsymbol{\sigma}_1\cdot \boldsymbol{\sigma}_2) +
\frac{g_4^4}{9 f_\pi^4}\left[- V_{BT}(r) + V_{CT}(r)\right]S_{12}.
\end{eqnarray}
Putting these contributions together, we finally obtain the
$\Lambda_c^+\bar{\Lambda}_c^+$ potential
\begin{equation}
V_{\Lambda_c^+\bar{\Lambda}_c^+} (r) =
V_{1\Lambda_c^+\bar{\Lambda}_c^+} (r) +
V_{2\Lambda_c^+\bar{\Lambda}_c^+} (r). \label{lamlam}
\end{equation}


\subsection{$\Sigma_c^0$-$\bar{\Sigma}_c^0$ potential}

In the $\Sigma_c^0$-$\bar{\Sigma}_c^0$ interaction, both one-pion-exchange
and two-pion-exchange are allowed. In the one-pion-exchange case, the
Lagrangian of the $\Sigma_c^0-\pi-\Sigma_c^0$ interaction can be written as
\begin{eqnarray}
\mathcal{L}_{\Sigma_c^0-\pi-\Sigma_c^0} &=& -
\frac{g_1}{\sqrt{2}f_\pi}~\bar{\Sigma_c}^{0}~\gamma^\mu\gamma_5~\partial_\mu\pi^{0,~-}
~\Sigma_c^{0,~+}.
\end{eqnarray}
The axial current interaction (one-pion-exchange) causes a spin-spin
potential
\begin{eqnarray}
V_{OPS} (r)(\boldsymbol{\sigma}_1\cdot \boldsymbol{\sigma}_2) =
-\frac{g_1^2}{3f_\pi^2}\left[I''(m\;,r) + \frac{1}{r}
I'(m\;,r)\right](\boldsymbol{\sigma}_1\cdot
\boldsymbol{\sigma}_2)\;, \end{eqnarray}
and a tensor potential
\begin{eqnarray}
 V_{OPT}(r) S_{12}= -\frac{g_1^2}{3f_\pi^2}\left[I''(m\;,r) - \frac{1}{r}
I'(m\;,r)\right]S_{12}\;,
\end{eqnarray}
where the function $I(m\;,r)$ is given in the Appendix of
Ref.{\cite{chen}}. Then we have one-pion-exchange caused potential
\begin{eqnarray}
V_{1\Sigma^0_c\bar{\Sigma}^0_c}(r) &=&
  V_{OPS}(r)(\boldsymbol{\sigma}_1 \cdot
\boldsymbol{\sigma}_2) + V_{OPT}(r) S_{12}
\end{eqnarray}

In the two-pion-exchange case, both spin-$\frac{1}{2}$
$\Sigma_c^0$ and $\Lambda_c^+$ can be the intermediate state. So, we have the
spin-$\frac{1}{2}$ intermediate state caused potential
\begin{eqnarray}
V_{2\Sigma^0_c\bar{\Sigma}^0_c}(r)
&=&\left[\left(\frac{g_1^4}{4f_\pi^4} + \frac{g_2^4}{f_\pi^4}\right)
(V_{BC}(r) + V_{CC}(r)) + \left(\frac{g_1^2}{2f_\pi^4} +
\frac{g_2^2}{f_\pi^4}\right)V_{TC}(r) +
\frac{1}{f_\pi^4}V_{2\pi-loop}(r)\right]\nonumber\\
 & + &\left(\frac{g_1^4}{4f_\pi^4} +
  \frac{g_2^4}{f_\pi^4}\right)\left[ V_{BS}(r) + V_{CS}(r)
  \right] (\boldsymbol{\sigma}_1 \cdot
\boldsymbol{\sigma}_2)\nonumber\\
& +& \left(\frac{g_1^4}{4f_\pi^4} +
\frac{g_2^4}{f_\pi^4}\right)\left[V_{BT}(r) + V_{CT}(r)\right]
S_{12},
\end{eqnarray}
where $g_1$ stands for the coupling constant in the case where
spin-$\frac{1}{2}$ $\Sigma_c^0$ is an intermediate state.

Moreover, spin-$\frac{3}{2}$ $\Sigma^*_c$, as an intermediate state,
would also contributes. The Lagrangian of the $\Sigma_c^{0*}-\pi-\Sigma_c^0$ interaction reads
\begin{eqnarray}
\mathcal{L}_{\Sigma_c^{0*}-\pi-\Sigma_c^0} &=& -
\frac{g_3}{\sqrt{2}f_\pi}\bar{\Sigma}^{0*\mu}\partial_\mu\pi^0\Sigma^0\;,
\end{eqnarray}
with $g_3$ being the coupling constant. Based on this Lagrangian,
following the same procedure used above, we have the
spin-$\frac{3}{2}$ $\Sigma^*_c$, as an intermediate state, caused potential
\begin{eqnarray}
V_{3\Sigma^0_c\bar{\Sigma}^0_c}(r) &=&
\frac{g_3^4}{9f_\pi^4}(V_{BC}(r) + V_{CC}(r)) +
\frac{g_3^2}{3f_\pi^4}V_{TC}
 +\frac{g_3^4}{36f_\pi^4}\left[- V_{BS}(r) + V_{CS}(r)\right]
(\boldsymbol{\sigma}_1\cdot \boldsymbol{\sigma}_2)\nonumber\\ &+&
\frac{g_3^4}{36f_\pi^4}\left[- V_{BT}(r) + V_{CT}(r)\right] S_{12}
\end{eqnarray}
Finally, we obtain the $\Sigma_c^0$-$\bar{\Sigma}_c^0$ potential
\begin{equation}
V_{\Sigma_c^0-\bar{\Sigma}_c^0} (r) =
V_{1\Sigma^0_c\bar{\Sigma}^0_c}(r) +
V_{2\Sigma^0_c\bar{\Sigma}^0_c}(r)+
V_{3\Sigma^0_c\bar{\Sigma}^0_c}(r)\label{sigsig}
\end{equation}

\subsection{$\Lambda_b^0$-$\bar{\Lambda}_b^0$ potential}
The same formulas can also be applied to the $\Lambda_b^0$-$\bar{\Lambda}_b^0$
interaction except the $c$-flavored heavy baryon (antibaryon) is replaced
by the $b$-flavored heavy baryon (antibaryon).

\section{Numerical Result and Discussion}

In the numerical calculation, we take $m_\pi = 0.135$GeV and
$f_\pi=0.132$GeV. We also choose the cutoff parameter $\Lambda=0.6-1.0$ GeV,
because in the chiral
perturbation theory, the momentum transfer is usually less than $1.0$GeV.

\begin{figure}[htbp]
\begin{center}
\scalebox{0.50}{\includegraphics{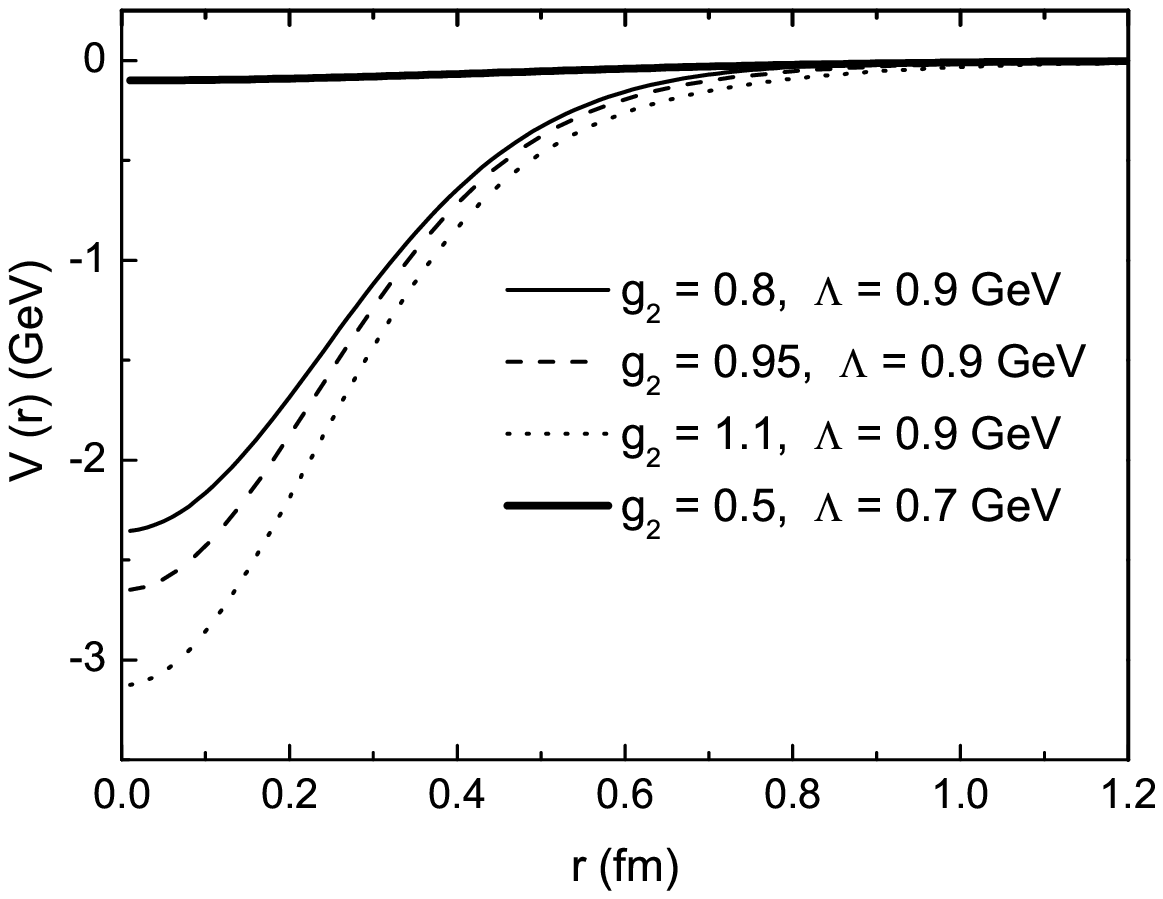}}
\scalebox{0.50}{\includegraphics{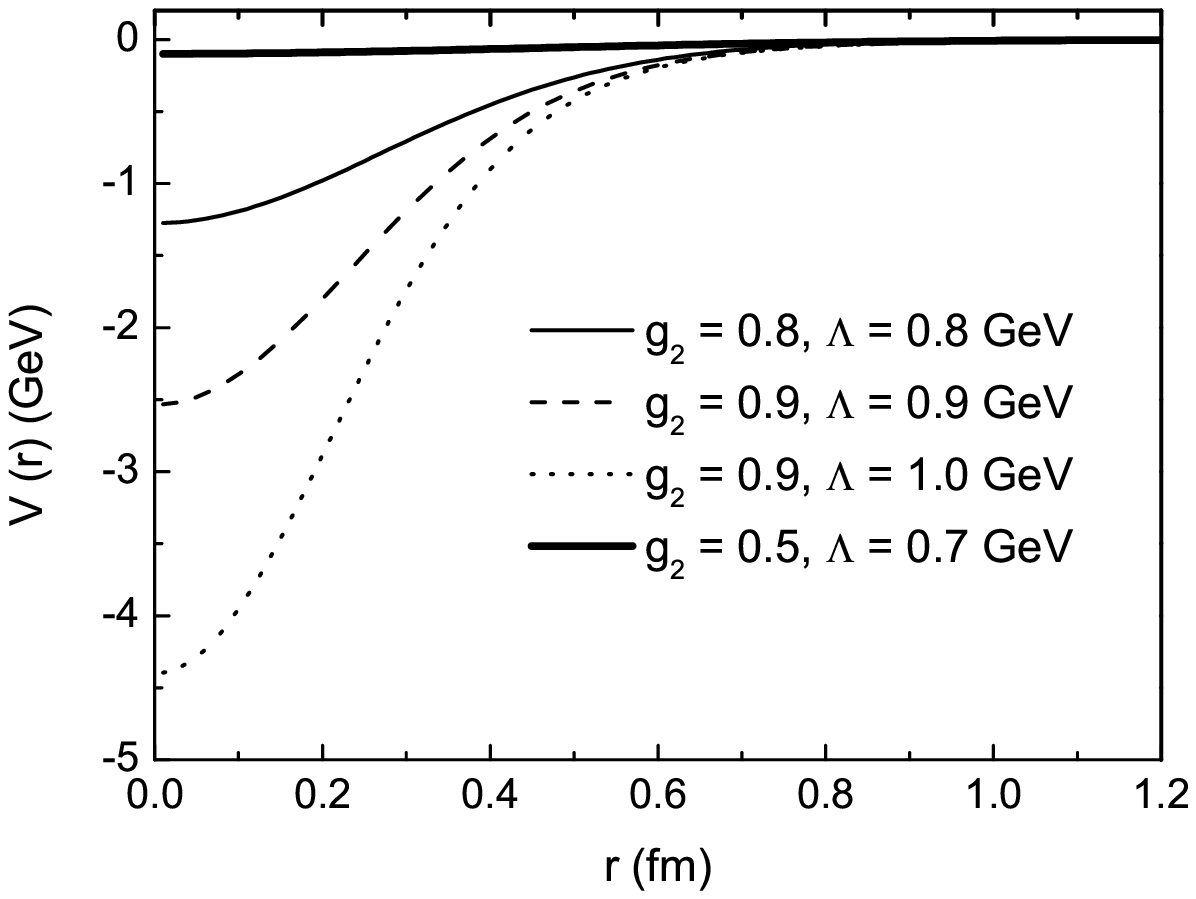}}
 \caption{The $\Lambda_c$-$\bar{\Lambda}_c$ potential
  in the singlet state with different $g_2$ but fixed $\Lambda$ (left figure) and
  different $\Lambda$ but fixed $g_2$ (right figure).}
  \label{singlet}
  \end{center}
\end{figure}

\begin{figure}[htbp]
\begin{center}
\scalebox{0.50}{\includegraphics{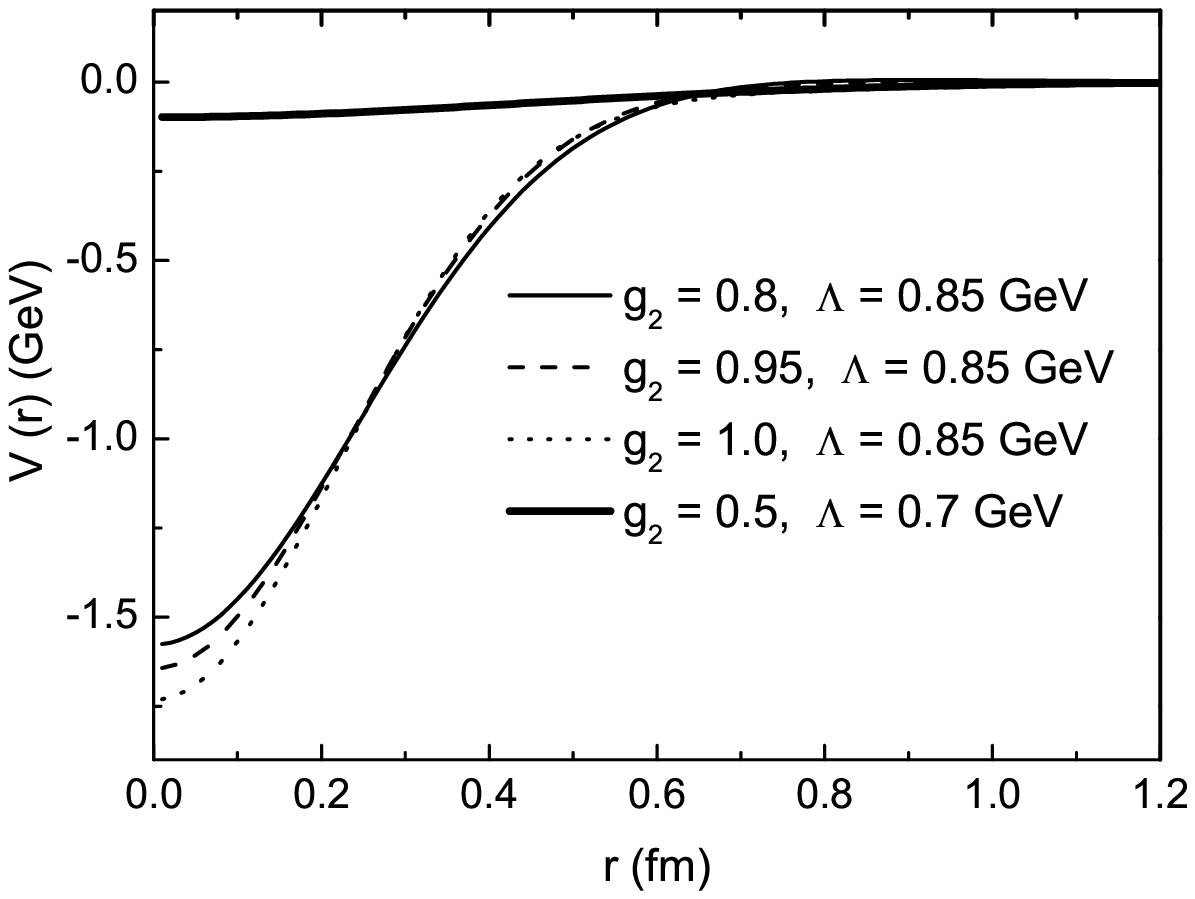}}
\scalebox{0.50}{\includegraphics{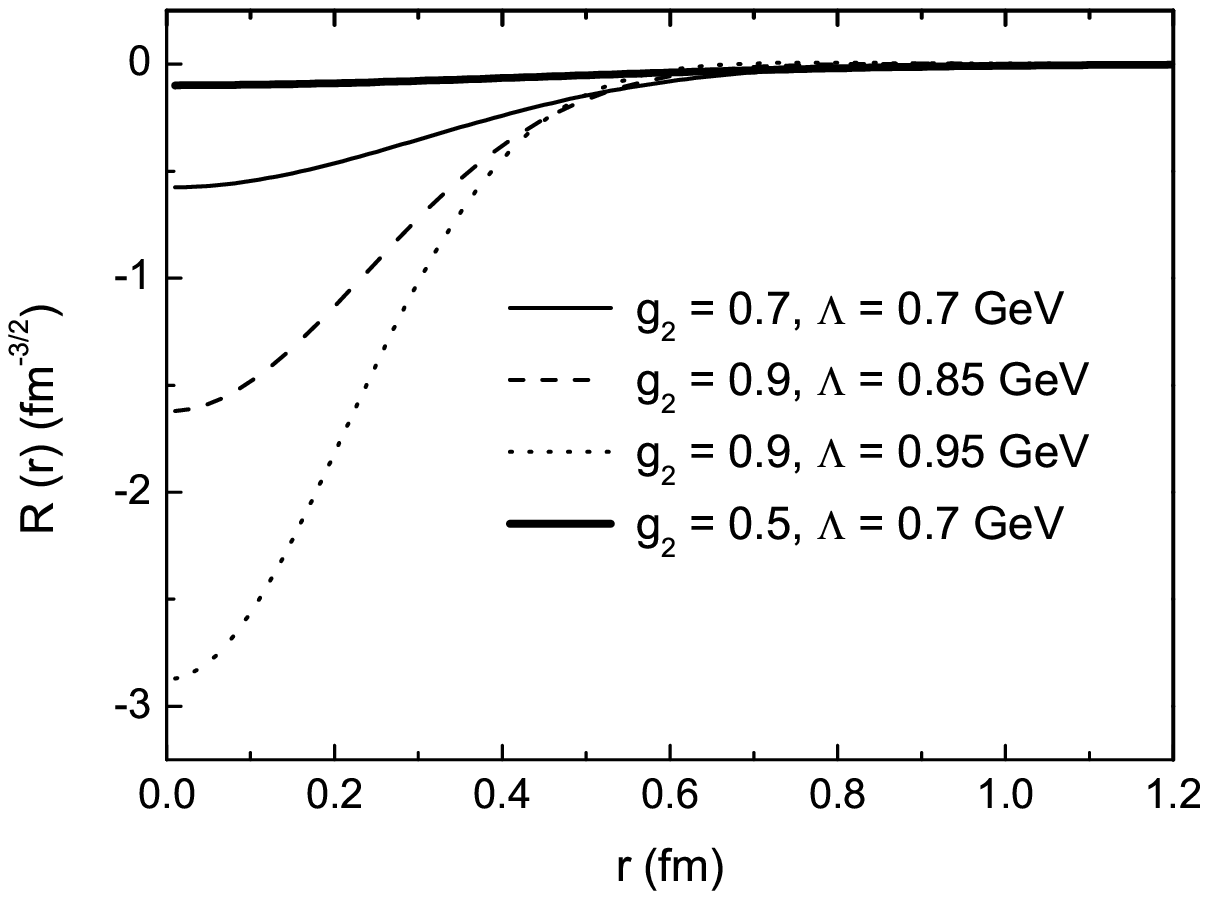}}
 \caption{The $\Lambda_c$-$\bar{\Lambda}_c$ potential
  in the triplet state with different $g_2$ but fixed $\Lambda$ (left figure) and
  different $\Lambda$ but fixed $g_2$ (right figure).}
  \label{triplet}
  \end{center}
\end{figure}

In the $\Lambda^+_c$-$\bar{\Lambda}^{+}_c$ system, the averaged mass
difference between $\Sigma_c^*$  and $\Lambda_c$ is about
$\Delta=0.234$GeV. The resultant potentials for the spin-singlet and
spin-triplet $\Lambda_c$-$\bar{\Lambda}_c$ states are plotted in
Figs.(\ref{singlet}) and (\ref{triplet}), respectively.
From these figures, we see that comparing with our previous result \cite{chen},
no matter in which states, the singular behavior of the potential
around origin is  greatly reduced. This indicates that the contribution from
large $\lambda$ values  is also important in the two-pion exchange
process. Moreover, the potentials become   more attractive with
increasing values of $g_2$ and $\Lambda$. This is reasonable,
because the larger $g_2$ value provides stronger coupling and
consequently stronger   potential. And the value of the cut-off
$\Lambda$ largely affects the depth of the   potential, the smaller
value of $\Lambda$ makes the shorter distance interaction
even more suppressed. It partly prevents the $\Lambda_c$ and
$\bar{\Lambda}_c$   getting too close, thus matches our treatment
of omitting the $s$-channel interaction. The line shape of these
potentials also tells us that the interaction between $\Lambda_c$
and $\bar{\Lambda}_c$ is attractive and might bind these particles together.

With these potentials, we can study the $\Lambda_c$-$\bar{\Lambda}_c$
scattering property. The partial wave Schr\"{o}dinger equation that the
$\Lambda_c$-$\bar{\Lambda}_c$ scattering obeys reads
\begin{equation}
\frac{d^2 u_l(r)}{dr^2} + \left[k^2 - \frac{l(l+1)}{r^2} - U(r)\right]u_l(r)=0,
\label{partsch}
\end{equation}
with the boundary condition
%
\begin{equation}
u_l(r)= krj_l(kr) + \int_0^\infty G_l(r,r')U(r')u_l(r')dr'
\end{equation}
where $G_l(r,r')$ is the Green function in the form of
\begin{eqnarray}
G_l(r,r')&  = & krr'j_l(kr)n_l(kr'),~~~~~~~~~~~~~~ r \leq r' \nonumber \\
   &  = & krr'j_l(kr')n_l(kr),~~~~~~~~~~~~~~ r \geq r'
\end{eqnarray}
and $U(r) = 2\mu V(r)$, $j_l(kr)$ and $n_l(kr)$ are the spherical Bessel function
and the spherical Neumann function, respectively \cite{stern}.
Then, the scattering phase shift $\delta_l(k)$ and the potential $V(r)$ has the relation
\begin{equation}
\tan \delta_l (k) = -\int_0^\infty r'j_l (kr')U(r')u_l(r')dr'.
\end{equation}

\begin{figure}[htbp]
\begin{center}
\scalebox{0.5}{\includegraphics{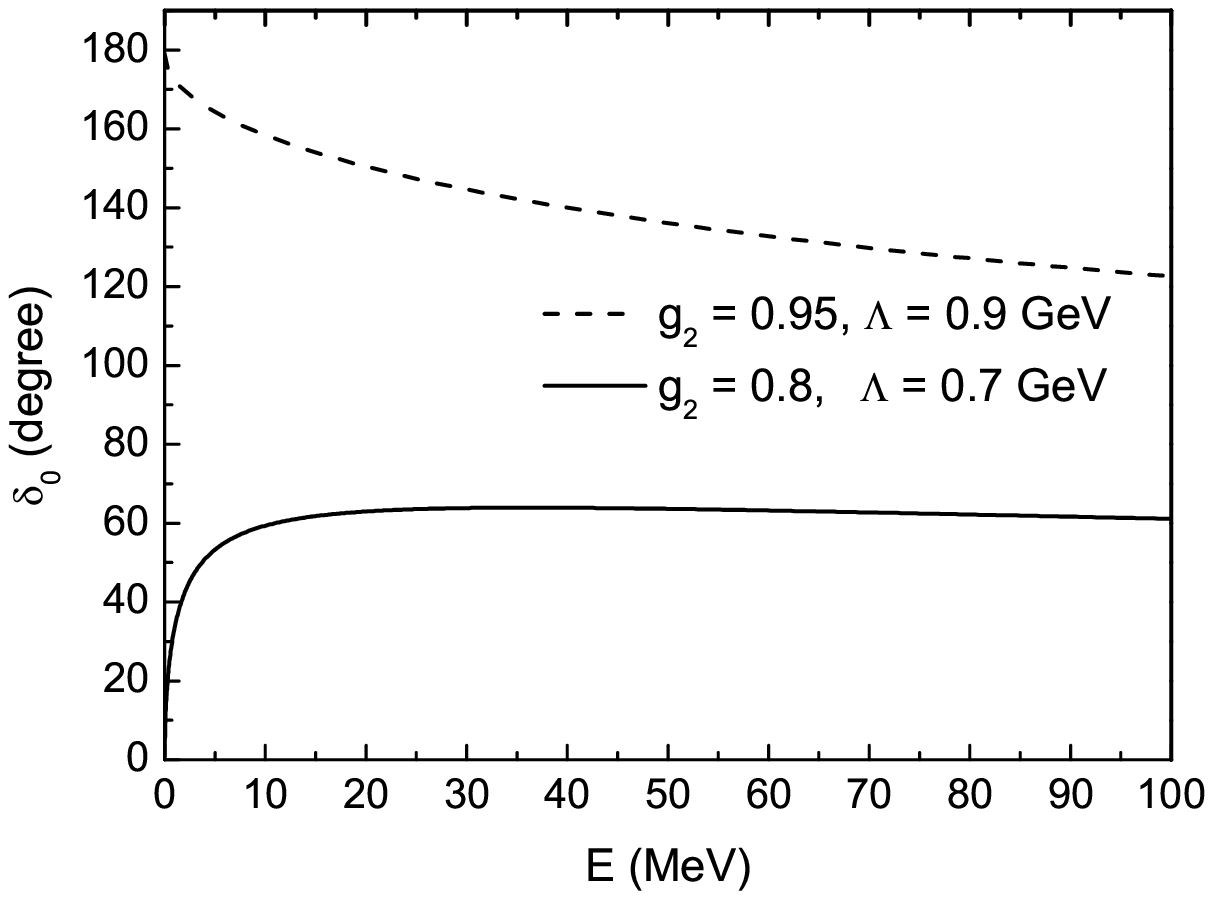}}%
\scalebox{0.5}{\includegraphics{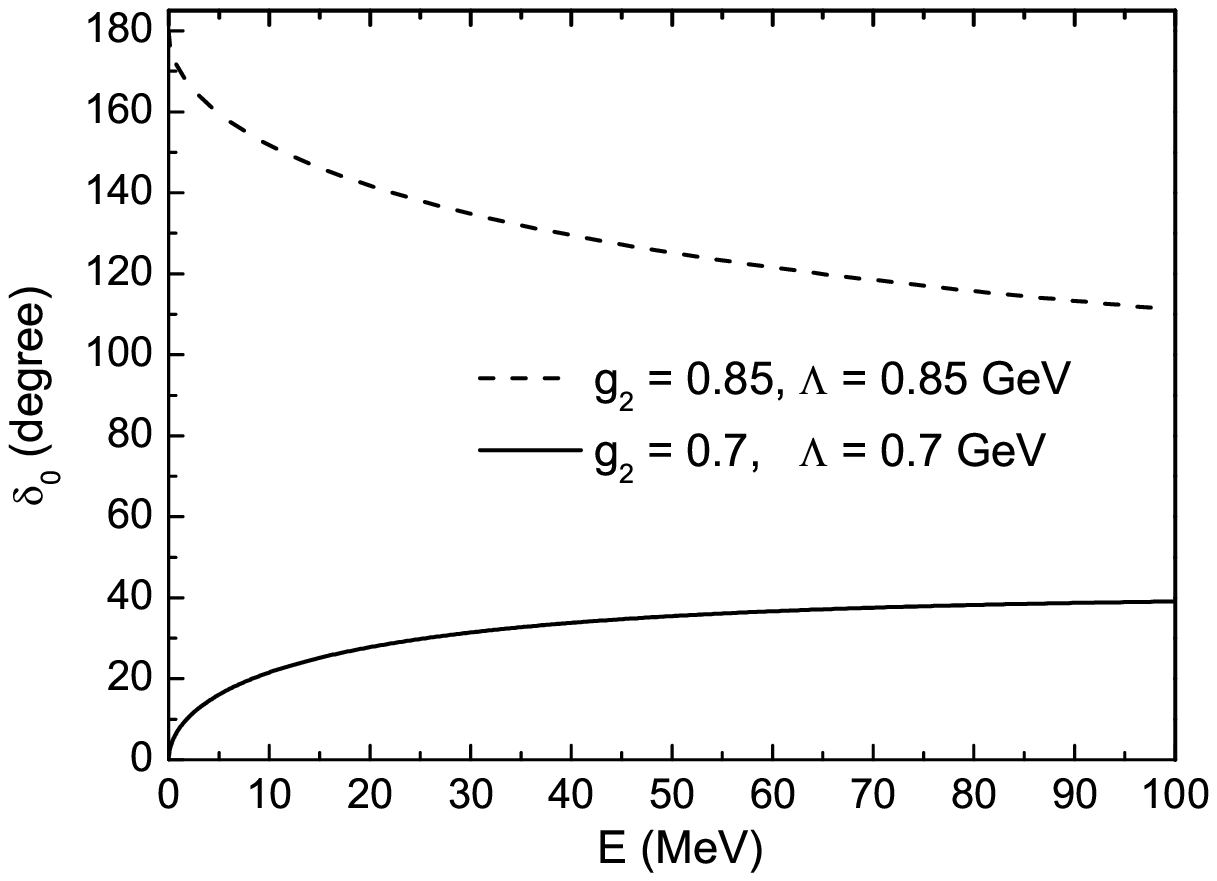}}%
 \caption{Phase shifts of the $\Lambda_c$-$\bar{\Lambda}_c$ system
 in the spin-singlet (left figure) and the spin-triplet (right figure) states. }
  \label{phase1}
  \end{center}
\end{figure}

Solving above equations numerically, we obtain scattering phase
shifts and plot them in Fig.\ref{phase1}.
From this figure, one sees that although all the potentials are attractive,
in some cases, the potential does not support a binding character, especially
in the case where the value of $g_2$ is extracted from the
$\Sigma_c \to \Lambda_c \pi$ decay data (the thick solid curve).
It means that the $\Lambda_c - \bar{\Lambda}_c$ could be bound
(the dashed curve) only when the coupling constant $g_2$ takes
a value larger than that from the data fitting, and the cut-off
$\Lambda$ is larger than that in the light baryon sector.

Moreover, we can also calculate the scattering length for concerned states by
\begin{equation}
a = -\lim_{k\rightarrow 0} \frac{\tan \delta_0(k)}{k}.
\end{equation}
The results are tabulated in Table \ref{scatteringlength}.
\begin{center}
\begin{table}[htbp]
\caption{\small Scattering length for the spin-singlet and
spin-triplet states in the $\Lambda_c$-$\bar{\Lambda_c}$ and
$\Sigma_c$-$\bar{\Sigma_c}$systems. } \centering
\begin{tabular}{c c  || c c }
\hline\hline
\multicolumn{2}{c ||}{$S=0$ state for $\Lambda_c$-$\bar{\Lambda}_c$} &
\multicolumn{2}{ c}{$S=1$ state for $\Lambda_c$-$\bar{\Lambda}_c$}\\
\hline
$g_2=0.95$, $\Lambda=0.9GeV$ &$a = 3.5fm$ &$g_2=0.85$, $\Lambda=0.85GeV$ &$a = 2.1fm$     \\
$g_2=0.8$, $\Lambda=0.7GeV$ &$a= -2.7fm$&$g_2=0.7$,
$\Lambda=0.7GeV$ &$a = -0.6fm$
\\\hline\hline \multicolumn{2}{c ||}{$S=0$ state for
$\Sigma_c$-$\bar{\Sigma}_c$} & \multicolumn{2}{ c}{$S=1$ state
$\Sigma$-$\bar{\Sigma}_c$}\\\hline
$g_1=0.85, \Lambda=1.1GeV$ &$a=3.5fm$ &$g_1=0.95, \Lambda=1.1GeV$ &$a=5.7fm$     \\
$g_1=0.8, \Lambda=0.8GeV$ &$a=-1.7fm$&$g_1=0.8, \Lambda=0.8GeV$ &$a=-0.8fm$   \\
\hline\hline
\end{tabular}
\label{scatteringlength}
\end{table}
\end{center}
The scattering lengths in this table also tell us that only those $g_2$ and
$\Lambda$ values, with which the attractive potential is much stronger
(denoted by dashed curve in Figs.\ref{singlet}-\ref{triplet}),
can produce an appropriate positive scattering length, which denotes a
bound $\Lambda_c$-$\bar{\Lambda_c}$ system, otherwise the system is unbound.

Based on the enlightenment from the scattering study, we further perform
the bound state calculation to check the condition for
forming a $\Lambda_c$-$\bar{\Lambda}_c$
bound state. Since we have the spin-spin interaction in
the $\Lambda_c$-$\bar{\Lambda}_c$ potential Eq.(\ref{lamlam}),
substituting such a potential into the Schr\"{o}dinger equation
and solving the equation numerically, we get the binding energy
for the spin splitted $S = 0$ and $S = 1$ states, respectively.
The results are tabulated in Table {\ref{tblambda}}.

\begin{center}
\begin{table}[htbp]
\caption{\small Binding energies (BE), as well as the masses of
heavy baryonium ($M_{\Lambda_c \bar{\Lambda}_c}$), in the
$\Lambda_c$-$\bar{\Lambda}_c$ system in various parameter cases. }
\centering
\begin{tabular}{c c c c || c c c c}
\hline\hline
\multicolumn{4}{c ||}{$S=0$ state} & \multicolumn{4}{ c}{$S=1$ state}\\
$|g_2|$ &$\Lambda(\mathrm{GeV})$& BE(MeV) & $M_{\Lambda_c
\bar{\Lambda}_c}$(GeV) & $|g_2|$ &$\Lambda(\mathrm{GeV})$& BE(MeV) &
$M_{\Lambda_c \bar{\Lambda}_c}$(GeV) \tabularnewline\hline
 $<$0.8 &$<$0.8& ----- & ----- & $<0.7$  &$<0.7$ & ----- & -----\\
0.9 &0.9&34  &4.538  & 0.9  &0.85  &75    &4.497    \\
0.9 &1.0&118  &4.45    & 0.9  &0.95  &285    &4.287 \\\hline
0.8 &0.9&3.25  &4.568    & 0.8  &0.85  &14    &4.558  \\
1.1 &0.9&166.2   &4.406   & 1.0  &0.85  &199   &4.373 \\
\hline\hline
\end{tabular}
\label{tblambda}
\end{table}
\end{center}
From this table, we see that with the extracted $g_2$ value of
$0.5\sim 0.57$ from the decay data, the
$\Lambda_c$-$\bar{\Lambda}_c$ system would not be bound. If we wish
$\Lambda_c$ and $\bar{\Lambda}_c$ being bound, no matter in the spin
singlet state or the spin triplet state, the coupling constant
should be much larger than the value extracted phenomenologically,
namely $g_2>0.8$ for the spin-singlet state and $g_2>0.78$ for the
spin-triplet state. This is in coincidence with those learned from
above scattering study. The result also shows the required ranges of
$g_2$ and $\Lambda$ for $\Lambda_c$-$\bar{\Lambda}_c$ binding:
$0.8<g_2\le 1.1$ and $0.8{\rm GeV}<\Lambda \le 1.0{\rm GeV} $ for
the spin singlet state and $0.7<g_2\le 1.0$ and $0.7{\rm GeV}<
\Lambda \le 0.95{\rm GeV} $ for the spin triplet state,
respectively. The mass of the corresponding baryonium is in the
region of $(4.406, \; 4.572]{\rm GeV}$ and  $(4.287, \; 4.572]{\rm
GeV}$ for the spin-singlet and spin-triplet states, respectively.

\begin{figure}[htbp]
\begin{center}
\scalebox{0.50}{\includegraphics{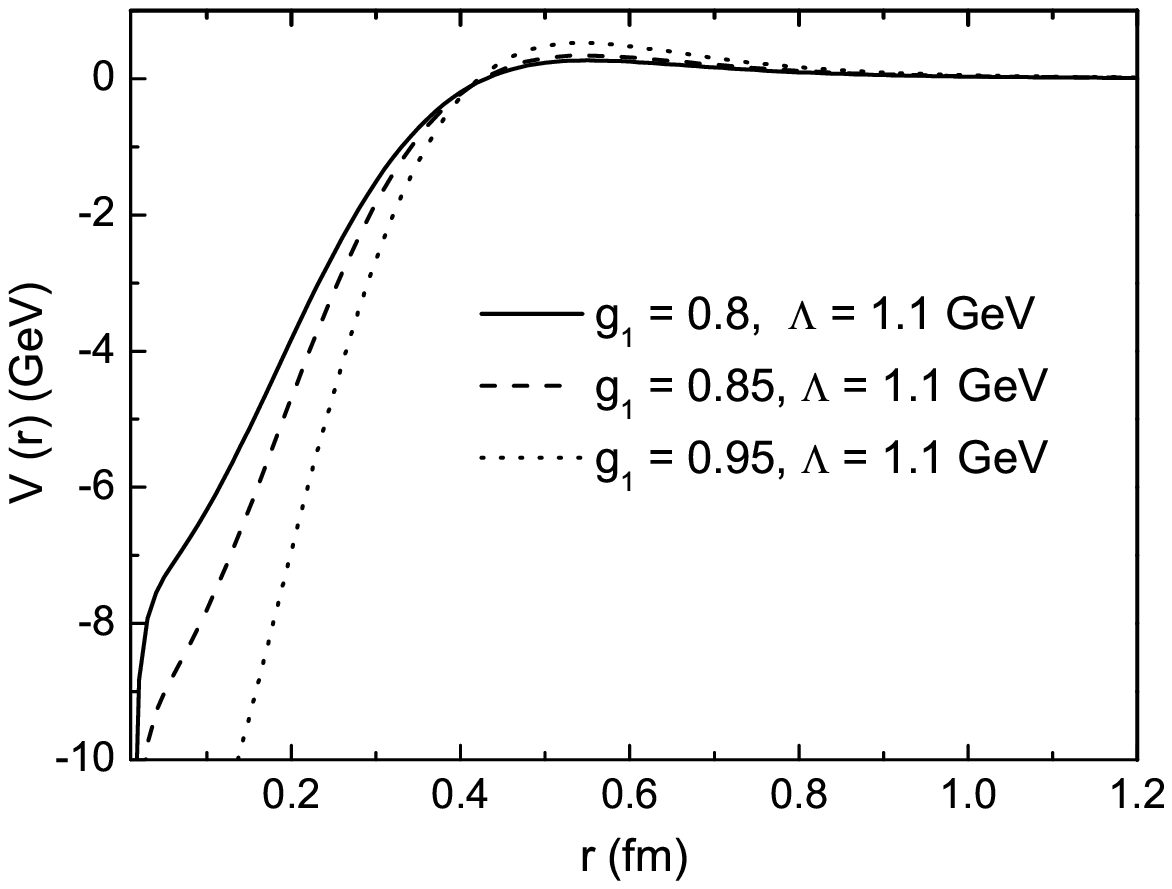}}%
\scalebox{0.50}{\includegraphics{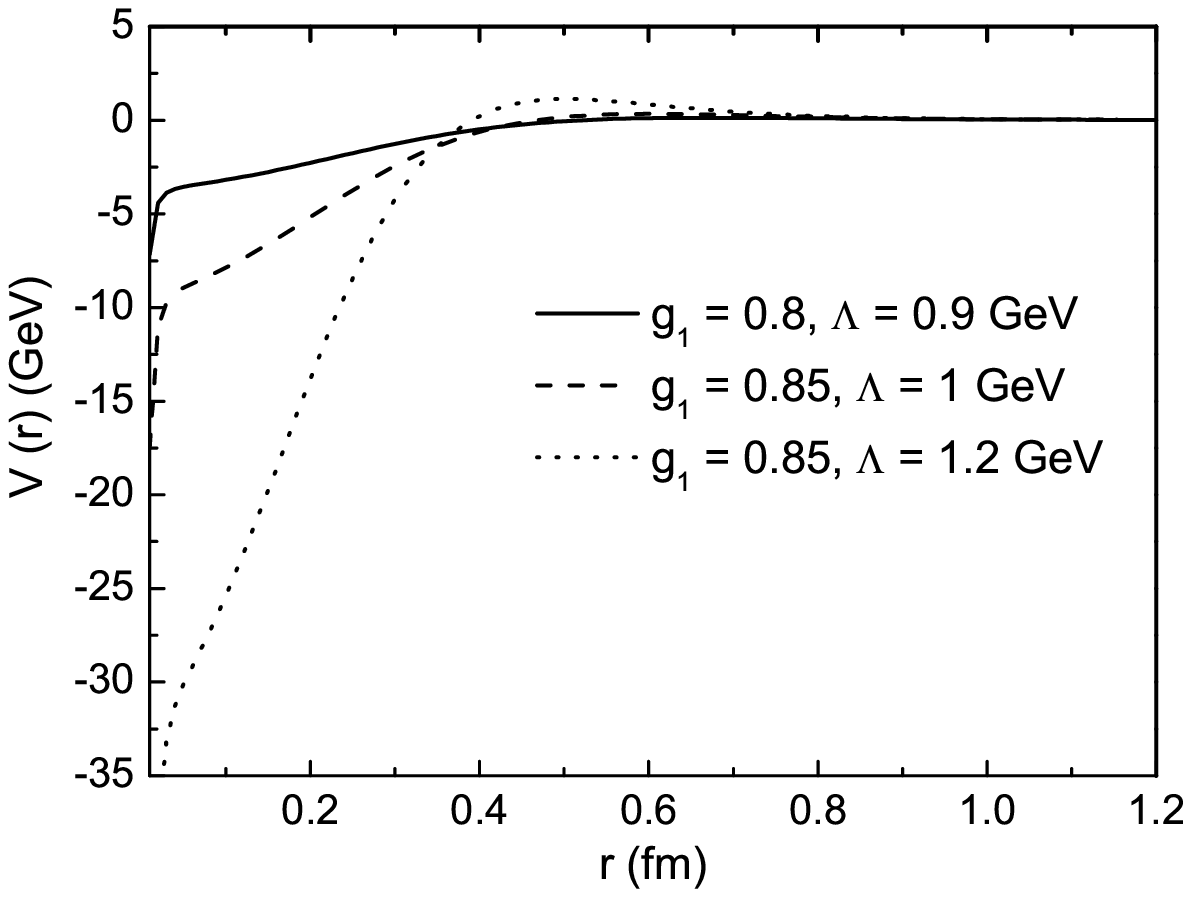}}%
 \caption{$\Sigma_c$-$\bar{\Sigma}_c$ potential
  in spin-singlet state.}
  \label{singlet2}
  \end{center}
\end{figure}

\begin{figure}[htbp]
\begin{center}
\scalebox{0.50}{\includegraphics{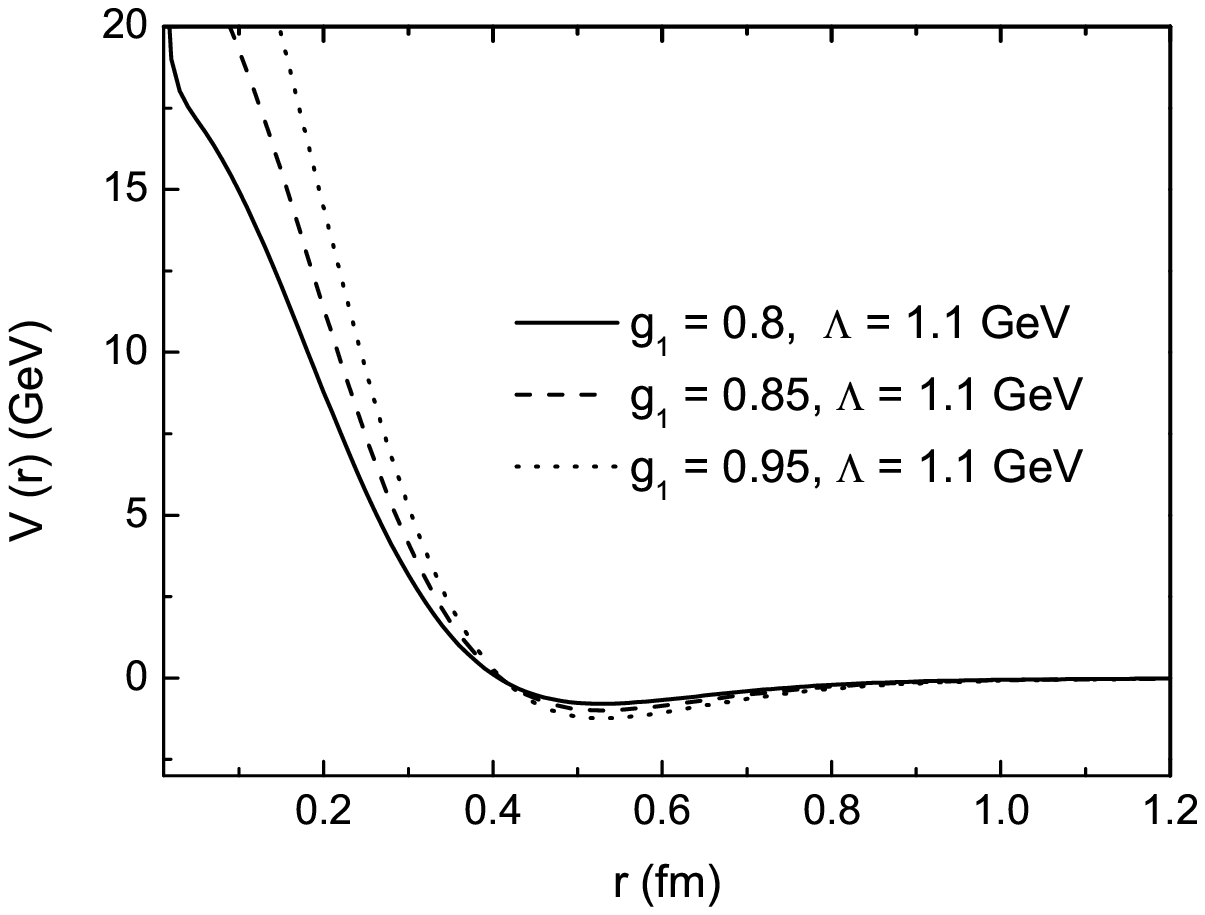}}%
\scalebox{0.50}{\includegraphics{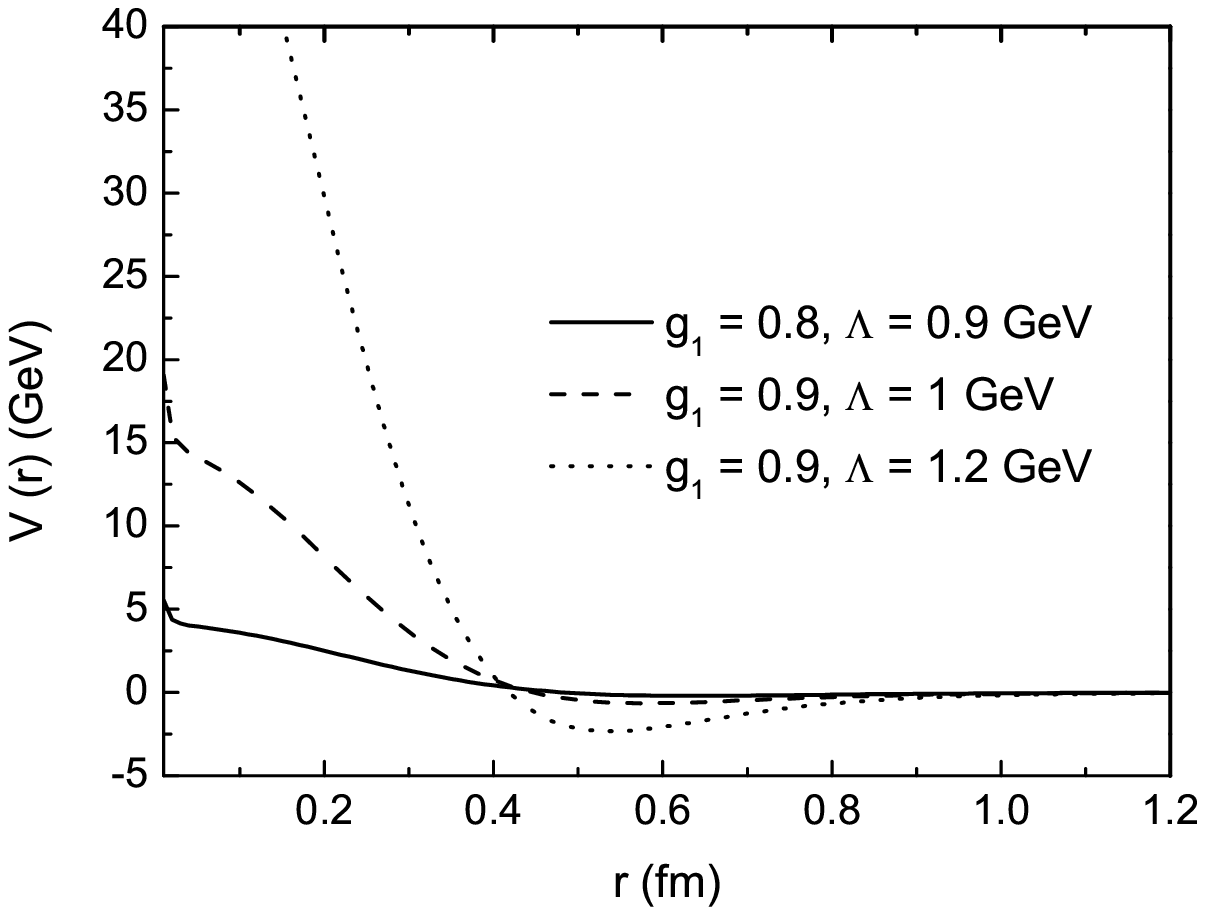}}%
 \caption{$\Sigma_c$-$\bar{\Sigma}_c$ potential
  in spin-triplet state.}
  \label{triplet2}
  \end{center}
\end{figure}

\begin{figure}[htbp]
\begin{center}
\scalebox{0.5}{\includegraphics{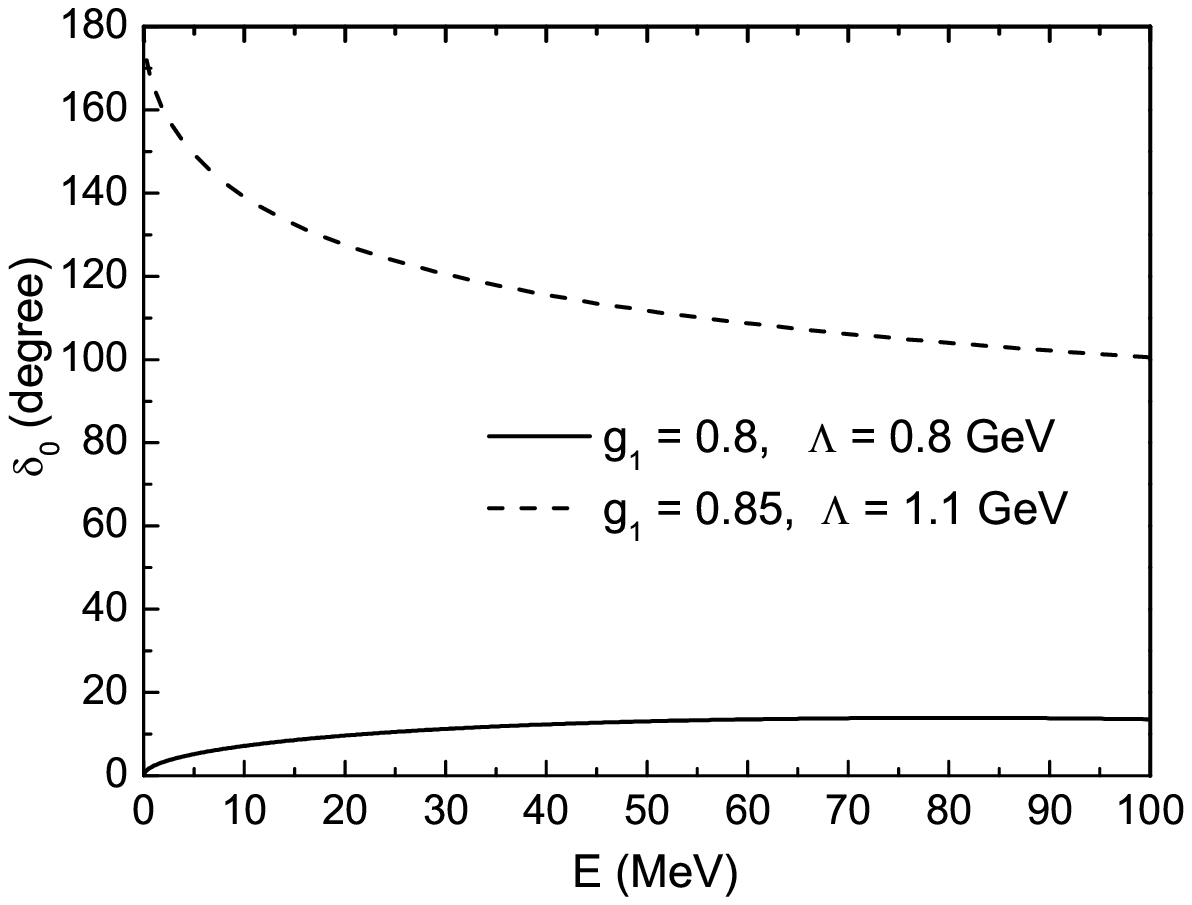}}%
\scalebox{0.5}{\includegraphics{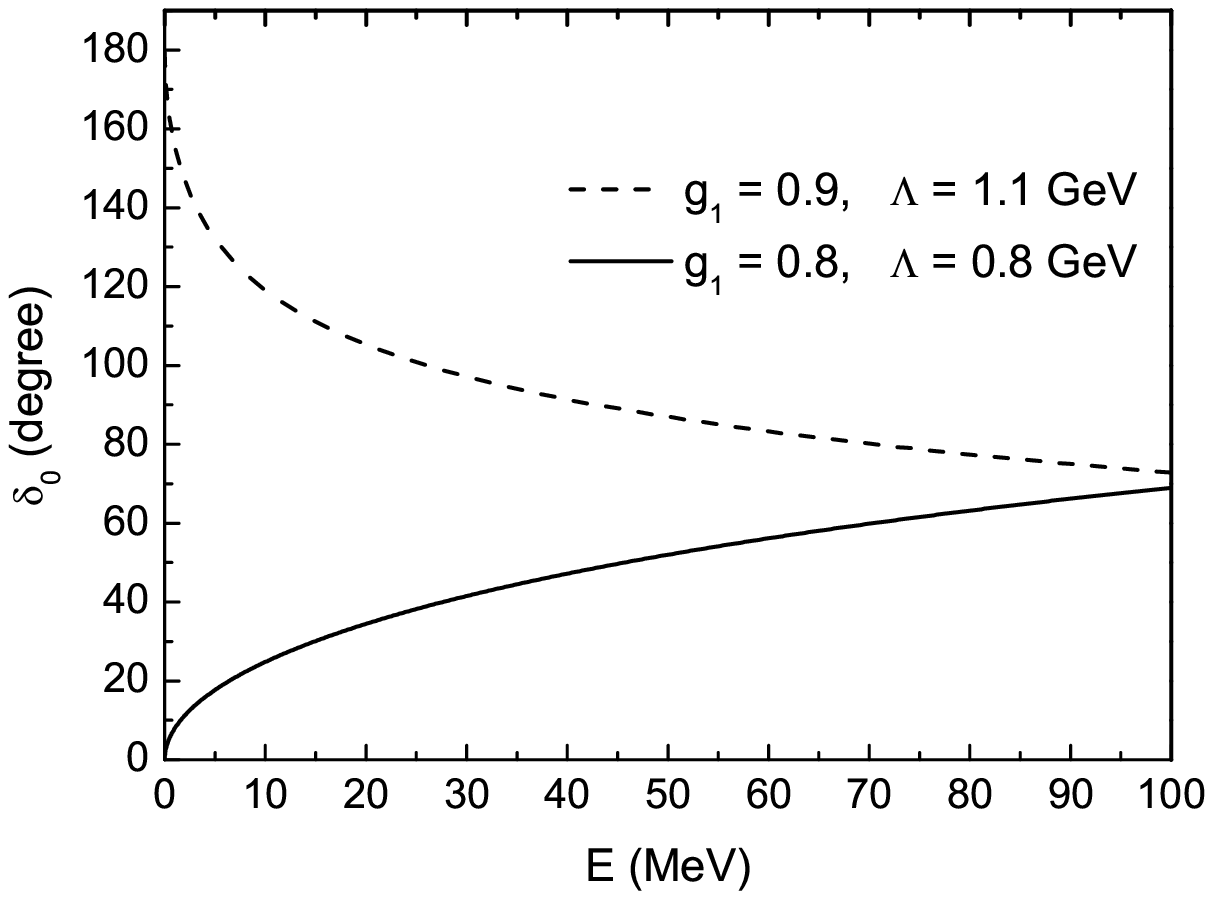}}%
 \caption{Phase shifts of the $\Sigma_c$-$\bar{\Sigma}_c$
 system in the spin-singlet (left figure)
 and the spin-triplet (right figure) states. }
  \label{phase2}
  \end{center}
\end{figure}

Same calculations can be performed for the $\Sigma_c$-$\bar{\Sigma}_c$ system as well.
The potentials for the spin-singlet and spin-triplet states are plotted in
Fig.\ref{singlet2} and Fig.\ref{triplet2}, respectively.
From these figures, it is shown that the potentials between
$\Sigma_c$ and $\bar{\Sigma}_c$ are different from those between
$\Lambda_c$ and $\bar{\Lambda}_c$, especially in the spin-triplet
state where the potential has a repulsive core in the short
distance. This is due to the contribution from the one-pion exchange,
which gives the spin-spin interaction, in the
$\Sigma_c$-$\bar{\Sigma}_c$ interaction. The phase shifts of the
$\Sigma_c$-$\bar{\Sigma}_c$ system
are plotted in Fig.\ref{phase2}.
Again, the system in some cases could be bound (dashed curve) and in the other cases
would be unbound (solid curve). However, due to lack of experimental data to fix
the $g_1$ value, it is necessary to examine the marginal condition for its binding.


Same as before, substituting the obtained $\Sigma_c$-$\bar{\Sigma}_c$
potential, Eq.(\ref{sigsig}), into the Schr\"{o}dinger equation
and solving it numerically, we have the binding character for the
$\Sigma_c$-$\bar{\Sigma}_c$ system.
The resultant binding energies for the $S = 0$ and $S = 1$ states
are tabulated in Table{\ref{tbsigma}}.

\begin{table}[htbp]
\caption{\small Binding energies (BE), as well as the masses of
heavy baryonium ($M_{\Sigma_c \bar{\Sigma}_c}$), for the
$\Sigma_c$-$\bar{\Sigma}_c$ system. }
\begin{center}
\begin{tabular}{c c c c || c c c c}
\hline\hline
\multicolumn{4}{c ||}{$S=0$ state} & \multicolumn{4}{ c}{$S=1$ state}\\
$|g_1|$ &$\Lambda(\mathrm{GeV})$& BE(MeV) & $M_{\Sigma_c
\bar{\Sigma}_c}$(GeV) & $|g_1|$ &$\Lambda(\mathrm{GeV})$& BE(MeV) &
$M_{\Sigma_c \bar{\Sigma}_c}$(GeV) \tabularnewline\hline
$<$0.8 &$<$0.95& ----- & ----- & $<0.8$  &$<0.95$ & ----- & -----\\
0.85 &1 &14  & 4.896  & 0.9  &1  &4.5    &4.9\\\hline
0.8 &1.1&21.7  &4.89    & 0.8  &1.1  &12.7    &4.897   \\
0.85 &1.1&29  & 4.88  & 0.85  &1.1  &39.9   &4.87   \\
\hline \hline
\end{tabular}
\end{center}
\label{tbsigma}
\end{table}
From this table, we find that as long as $g_1>0.8$ and
$\Lambda>1.0$GeV in the spin-singlet state and $g_1>0.9$ and
$\Lambda>1.0$GeV in the spin-triplet state, the
$\Sigma_c$-$\bar{\Sigma}_c$ system could be bound. And also the
spin-triplet state is slightly easier to be bound than the
spin-triplet state. The result also shows the required ranges of
$g_1$ and $\Lambda$ for the $\Sigma_c$-$\bar{\Sigma}_c$ binding:
$0.8<g_2\le 0.85$ and $0.95{\rm GeV}<\Lambda \le 1.1{\rm GeV} $ for
the spin singlet state and $0.8<g_2\le 0.9$ and $0.95{\rm
GeV}<\Lambda \le 1.1{\rm GeV} $ for the spin triplet state,
respectively. The mass of the corresponding baryonium is in the
region of $(4.880, \; 4.910]$ and $(4.870, \; 4.910]{\rm GeV}$ for
the spin-singlet and spin-triplet states, respectively.

The similar study can be done for the systems with the bottom
flavor. In the $\Lambda^+_b$-$\bar{\Lambda}^{+}_b$ system, the
averaged mass difference between $\Sigma_b^*$  and $\Lambda_b$ is
about $\Delta=0.114$GeV. With the same reason in the charm flavor sector,
namely due to lack of the experimental data to fix the $g_b$ value,
we also examine the marginal condition for its binding. Carrying out the
same procedure, we obtain the binding character of the
$\Lambda_b$-$\bar{\Lambda}_b$ system. The resultant binding
energies in the $S = 0$ and $S = 1$ states are tabulated in
Table{\ref{tblambdab}}.
\begin{table}[htb]
\caption{\small Binding energies (BE), as well as the masses of
heavy baryonium ($M_{\Lambda_b \bar{\Lambda}_b}$), in the
$\Lambda_b$-$\bar{\Lambda}_b$ system. }
\begin{center}
\begin{tabular}{c c c c || c c c c}
\hline\hline
\multicolumn{4}{c ||}{$S=0$ state} & \multicolumn{4}{ c}{$S=1$ state}\\
$|g_b|$ &$\Lambda(\mathrm{GeV})$& BE(MeV) & $M_{\Lambda_b
\bar{\Lambda}_b}$(GeV) & $|g_b|$ &$\Lambda(\mathrm{GeV})$& BE(MeV) &
$M_{\Lambda_b \bar{\Lambda}_b}$(GeV) \tabularnewline\hline
 $<$0.65 &$<$0.8& ----- & ----- & $<0.55$  &$<0.85$ & ----- & -----\\
0.8 &0.85 &25.5  & 11.21   & 0.6  &0.85  &4.1    &11.23    \\\hline
0.68 &0.8&7.5  &11.23     & 0.55  &0.9  &8    &11.23   \\
0.7 &0.9&15.6  & 11.22   & 0.6  &0.9  &38.5    &11.2
\\\hline\hline
\end{tabular}
\end{center}
\label{tblambdab}
\end{table}
The result shows that for the $\Lambda_b$-$\bar{\Lambda}_b$ system,
a relatively smaller $g_b$ value can make the spin-triplet state
bound. The result also presents the required ranges of $g_b$ and
$\Lambda$ for the $\Lambda_b$-$\bar{\Lambda}_b$ binding:
$0.65<g_b\le 0.8$ and $0.8{\rm GeV}\le \Lambda \le 0.9{\rm GeV} $
for the spin-singlet state and $0.55 < g_b\le 0.6$ and $0.85{\rm
GeV} \le \Lambda \le 0.9{\rm GeV} $ for the spin-triplet state,
respectively. The mass of the corresponding baryonium is in the
region of $(11.21, \; 11.24]$ and $(11.2, \; 11.24]{\rm GeV}$ for
the spin-singlet and spin-triplet states, respectively.

We also notice that someone has calculated the value of
$g_b$ \cite{stefan1,stefan2}, recently. They give $|g_b| = 0.475 \pm
0.050$ for the $\Sigma_b^*$-$\pi$-$\Lambda_b$ coupling. This value seems
too small to support a bound $\Lambda_b$-$\bar{\Lambda}_b$ state. However, the final
conclusion should not be made before some issues are clarified,
like, whether or not the $\Sigma_b^*$-$\pi$-$\Lambda_b$ coupling can be
straightforwardly applied to the loop calculation. When future
decay data of the $b$-flavored baryon become available, we would be able to extract
a physical value of $g_b$. If the extracted $g_b$ is consistent with
the marginal $g_b$ value for binding in this calculation, one might
confirm such a $b$-flavored heavy-baryonium.


\section{Conclusion}

The heavy-baryon-anti-heavy-baryon systems are studied in the
framework of heavy baryon chiral perturbation theory. The
potentials for the $\Lambda_c$-$\bar{\Lambda}_c$,
$\Sigma_c$-$\bar{\Sigma}_c$ and $\Lambda_b$-$\bar{\Lambda}_b$
interactions are derived with the
two-$pion$ exchange mechanism. Unlike our previous work, we
use the holonomic potential to investigate the scattering and
binding characters in this paper.
The scattering characters of these systems are calculated by
solving the partial Schr\"{o}dinger equation. From the obtained
phase shifts and the scattering lengths, it is found that the
$\Lambda_c$-$\bar{\Lambda}_c$ system could be bound with a $g_2$
value larger than that extracted phenomenologically from the decay
data of charmed baryons or estimated by Ref.\cite{T.M.Yan}.
For the $\Sigma_c$-$\bar{\Sigma}_c$ system, since we do not
have available decay data to fix $g_1$, whether the system is
bound depends on the selected value of $g_1$. To confirm these
results, the bound state calculations are further performed.
It is shown that marginal $g_2$ value for binding is about $0.8$
which is larger than the physical value of about $0.5\sim 0.57$.
In the $\Sigma_c$-$\bar{\Sigma}_c$ system, the marginal $g_1$
value for binding is also estimated. The minimum $g_1$ value is
about $0.85$. This value should be compared with that extracted
from the future data to affirm whether the $\Sigma_c$-$\bar{\Sigma}_c$
system could be bound. The $\Lambda_b$-$\bar{\Lambda}_b$ system is
studied as well. It is found that the minimum $g_b$ value for
binding is much smaller than that for the $\Lambda_c$-$\bar{\Lambda}_c$
system. If the $g_b$ value extracted from the future decay data of
the $b$-flavored baryon can meet this value,  one might confirm such
a $\Lambda_b$-$\bar{\Lambda}_b$ heavy baryonium.

It should be mentioned that above conclusions are also related the cutoff
value which is assumed to be similar to that for the light hadron sector
in chiral perturbation theory. A similar situation was met in
Ref.\cite{liuyr}, where a relative large cutoff is also required for
a possible molecule state in the $\Lambda_c$-$\Lambda_c$ system in
the one-pion-exchange model. Furthermore, it is worthwhile to
emphasize that in order to more realistically affirm whether the
heavy-baryon-anti-heavy-baryon system could have a bound state,
namely a heavy baryonium, the annihilation channel and couple channel
effects on the heavy baryonium potential should also be taken into account.
In particular, a study in the quark-gluon degree of freedom is necessary.

Even some corrections should be further considered, our results are much more
reliable and stable than those in our early calculation \cite{chen}. From the
regions of possible heavy baryonium masses, we conjecture that up to this stage,
$Y(4260)$ and $Y(4360)$ could be a spin-triplet $\Lambda_c$-$\bar{\Lambda}_c$
baryonium, but $Y(4660)$ could not be a $\Lambda_c$-$\bar{\Lambda}_c$ baryonium
in either spin-singlet or spin-triplet state, and $Y(10890)$ could not be a
$\Lambda_b$-$\bar{\Lambda}_b$ baryonium either. Moreover, because $Z^{\pm}(4430)$
is a electrically changed state and $Z_c(3900)$ is out of the possible binding
range, they are nothing to do with the $\Lambda_c$-$\bar{\Lambda}_c$ baryonium.

\vspace{0.0cm}

{\bf Acknowledgments}

This work was supported in part by the National Natural Science
Foundation of China(NSFC) under the grants 10935012, 10821063,
11175249, 11035006 and 11165005.

\newpage \vspace{.3cm}

\newpage
\vspace{0.5cm}


\end{document}